\documentclass[pdflatex,sn-mathphys-num]{sn-jnl}% Math and Physical Sciences Numbered Reference Style
\usepackage{graphicx}%
\usepackage{multirow}%
\usepackage{amsmath,amssymb,amsfonts}%
\usepackage{amsthm}%
\usepackage{mathrsfs}%
\usepackage[title]{appendix}%
\usepackage{xcolor}%
\usepackage{textcomp}%
\usepackage{manyfoot}%
\usepackage{booktabs}%
\usepackage{algorithm}%
\usepackage{algorithmicx}%
\usepackage{algpseudocode}%
\usepackage{listings}%

\usepackage{latexsym}

\usepackage{amsmath}
\usepackage{bm}
\usepackage{braket}
 \usepackage{amsfonts} 
 \usepackage[legacycolonsymbols]{mathtools}

\usepackage{color}
\theoremstyle{thmstyleone}%
\theoremstyle{thmstyletwo}%

\theoremstyle{thmstylethree}%

\begin{document}

\title[Article Title]{Violation of the method of images in non-Markovian processes and its connection to stochastic heat}

%%=============================================================%%
%% GivenName	-> \fnm{Joergen W.}
%% Particle	-> \spfx{van der} -> surname prefix
%% FamilyName	-> \sur{Ploeg}
%% Suffix	-> \sfx{IV}
%% \author*[1,2]{\fnm{Joergen W.} \spfx{van der} \sur{Ploeg} 
%%  \sfx{IV}}\email{iauthor@gmail.com}
%%=============================================================%%

\author*[]{\fnm{Takuya } \sur{Saito}}\email{tsaito@phys.aoyama.ac.jp}

\author[]{\fnm{Yuta} \sur{Sakamoto}}\email{sakamoto2021@phys.aoyama.ac.jp}
%% \equalcont{These authors contributed equally to this work.}

\author[]{\fnm{Takahiro} \sur{Sakaue}}\email{sakaue@phys.aoyama.ac.jp}
%% \equalcont{These authors contributed equally to this work.}

\affil*[1]{\orgdiv{Department of Physical Sciences}, \orgname{Aoyama Gakuin University}, \orgaddress{\street{Chuo-ku}, \city{Sagamihara}, \postcode{252-5258}, \state{Kanagawa}, \country{Japan}}}

% \affil[2]{\orgdiv{Department}, \orgname{Organization}, \orgaddress{\street{Street}, \city{City}, \postcode{10587}, \state{State}, \country{Country}}}

%% \affil[3]{\orgdiv{Department}, \orgname{Organization}, \orgaddress{\street{Street}, \city{City}, \postcode{610101}, \state{State}, \country{Country}}}

%%==================================%%
%% Sample for unstructured abstract %%
%%==================================%%

\def\Vec#1{\mbox{\boldmath $#1$}}
\def\degC{\kern-.2em\r{}\kern-.3em C}

\def\SimIneA{\hspace{0.3em}\raisebox{0.4ex}{$<$}\hspace{-0.75em}\raisebox{-.7ex}{$\sim$}\hspace{0.3em}} 

\def\SimIneB{\hspace{0.3em}\raisebox{0.4ex}{$>$}\hspace{-0.75em}\raisebox{-.7ex}{$\sim$}\hspace{0.3em}}

\abstract{This article discusses a failure of the widely used method of images to describe the time evolution of probability distributions in diffusive processes with memory. A walker in one-dimensional space draws a dead or a surviving path depending on whether it has touched a target during stochastic evolution. For the dead walker, we define its conjugate twin paired by paths spatially reflected at the first passage time. The probability distribution of the reflected dead path coincides with that of the free image walker in a physical domain, but not generally with that of the original dead walker for non-Markovian processes, which violates the method of images. For systems reducible to the generalized Langevin equation with the fluctuation-dissipation relation, we propose an energetic interpretation in terms of a path-memory force, where the path-probability ratio of the dead to the reflected path obeys an analogous relation to the fluctuation theorem associated with heat from the reflected to the original dead walker. This framework provides a quantitative basis as well as an intuitive picture of how and why the method of images breaks down for non-Markovian processes.}


%%\pacs[JEL Classification]{D8, H51}

%%\pacs[MSC Classification]{35A01, 65L10, 65L12, 65L20, 65L70}

\maketitle

\section{Introduction}

Calculation of electric potentials created by a point charge facing conducting planes is a classical subject in electrostatics~\cite{Jackson}. A similar situation arises, for instance, in the low Reynolds number hydrodynamics computing fluid flow under stick boundary conditions~\cite{Blake}. Solving such boundary problems is conveniently done thanks to a technique invented by Thomson in 1849~\cite{Thomson}, where one places a fictitious ``image" source to satisfy the boundary condition together with the real source. This technique, called the method of images (MIs), provides a versatile utility to various boundary problems. One appreciates that its applicability is rooted in the geometry, or more precisely, spatial symmetry of the system under consideration, which manifests in the Laplacian operator in the above examples.  
As such, the MIs work also for the dynamical phenomena, most notably, in the processes described by the diffusion equation, suggesting its use in analyzing the behaviors of random walkers~\cite{Redner,RevModPhys_Chndrasekhar_1943}.
Indeed, researchers have routinely leveraged it in the first passage problems, which often require
to know the time evolution of the probability distribution of relevant stochastic variables under absorbing boundary conditions~\cite{Redner,RevModPhys_Chndrasekhar_1943}.
Given its power and handiness, however, it is also important to figure out the limitations~\cite{AdvChemPhys_Chechkin_Metzler_2006,PRE_Kantor_Kardar_2007,PRE_Zoia_Kardar_2007,PRE_Amitai_2010,JCP_Sanders_2012,JCP_Forsling_2014,EPL_Jeon_2011,PRR_Sakamoto_Sakaue_2023}. Rather obviously,  the MIs cannot be used in systems with potential $U(x)$ that destroys the spatial symmetry. Even without potential, however, the MIs break down in systems with memory, i.e., the non-Markovian processes.
Here, the simple use of MIs has been known to lead to erroneous consequences~\cite{PRE_Amitai_2010,JCP_Sanders_2012,PRR_Sakamoto_Sakaue_2023}, but its underlying physics has not been clarified yet.

This article analyzes the MI structure of the path probability of non-Markovian Gaussian random walkers in the context of the first passage problem.
A remarkable point of the non-Markovian Gaussian processes is compatibility with the generalized Langevin equation (GLE).
In isothermal systems, we incorporate the fluctuation-dissipation relation (FDR) of the second kind that identifies ``friction" and  ``force balance",  from which ``memory force" is extracted to construct energetic quantities.
The main text is organized as follows:
We begin by revisiting the MI violation in Sec.~\ref{MI_revisit_sec}.
Making its conjugate twin of spatially reflected path for a specific first passage path, Sec.~\ref{FT_formalism_sec} demonstrates that their path probability ratio creates an analogous family member of the fluctuation theorem (FT)~\cite{Maes_2021,PRE_Crooks_2000,RPP_Seifert_2012,PRX_Jarzynski_2017,AdvPhys_Roldan_2023,JStatMech_Chernyak_Chertkov_Jarzynski_2006,JStatMech_GarciaGarcia_2012,PRL_Hatano_Sasa_2001,JPhysAMathGen_Speck_Seifert_2005}. Subsequently, Sec.~\ref{TD_sec} exhibits main results that the analogous FT formalism allows us to quantify the violation of the MIs in non-Markovian processes in terms of stochastic heat~\cite{Sekimoto_book}. 
Section~\ref{Discussion_sec} provides pertinent discussions and perspectives on memory force and heat flow.
In the end, Sec.~\ref{Conclusion_sec} concludes this study.
Supplemental information and some detailed calculations are included in the Appendix.

\section{Method of images revisited}
\label{MI_revisit_sec}

Consider a one-dimensional random walk in an unbounded domain.
Let $P_0(x,t|x_0)$ be the probability density function (PDF) of the position of random walkers with an initial location $x=x_0$. For Markovian walkers~\cite{vanKampen,Gardiner,PhysRep_Bouchaud_Georges_1990,Kubo_Toda_Hashitsume}, the Gaussian-distributed evolutions of $P_0(x,t|x_0)$ are found by the solution of diffusion equation $\partial_t P_0(x,t|x_0) = \partial_x^2 P_0(x,t|x_0)$ with initial condition $P_0(x,0|x_0)=\delta(x-x_0)$ and natural boundary condition, i.e., vanishing probability $P_0(\pm \infty,0|x_0)=0$ at infinity.

\begin{figure}[b]
	\centering
	\includegraphics[width=0.25\textwidth]{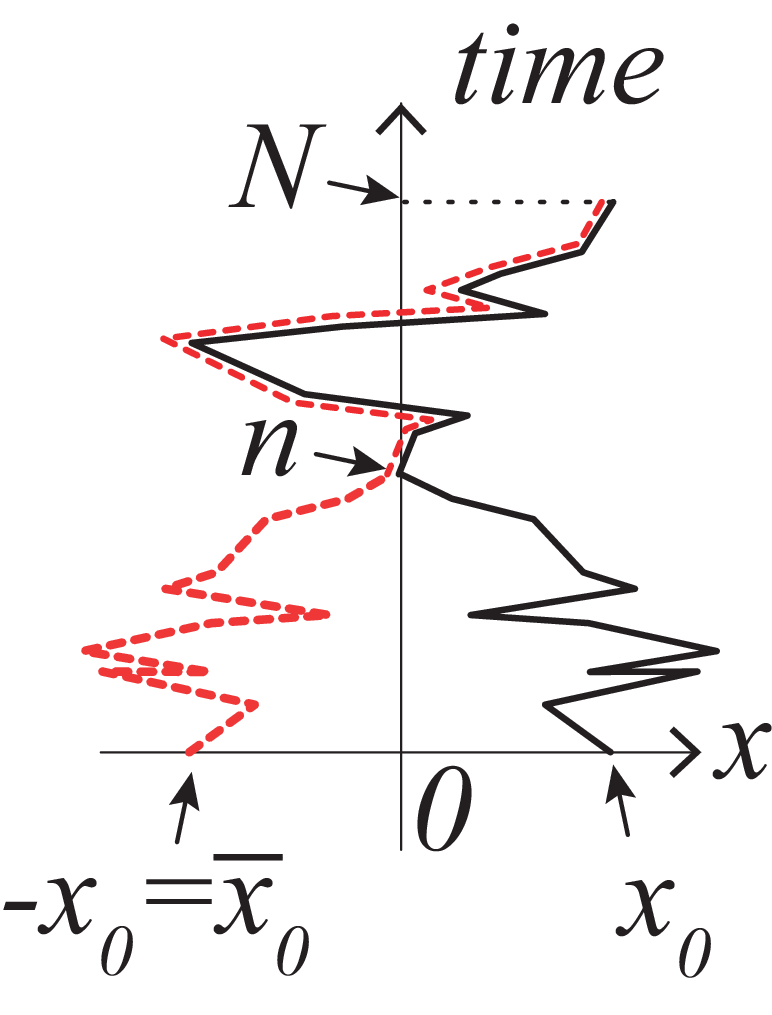}
	\caption{
	Schematic spatiotemporal plots of the original dead path (black solid) and the reflected dead path (red dashed) for the time period $[0,N]$ along the vertical time axis. 
	%% The reflected dead path is constructed by partially folding the original path $\{ x_i \}_0^n$ up to $n$ at $x=0$ on the spatiotemporal plane.
	}
	\label{fig1}
			\vspace{0.2 cm}
\end{figure}

In a first passage problem, throughout the article, an absorbing boundary is placed at $x=0$ as shown in Fig.~\ref{fig1}.
One then asks the time evolution of the survival PDF $P_S(x,t|x_0)$ for the walker starting from $x_0>0$, which obeys to $\partial_t P_S(x,t|x_0) = \partial_x^2 P_S(x,t|x_0)$ in the physical domain $x \ge 0$ with the initial $P_S(x,0|x_0)=\delta(x-x_0)$ and the boundary conditions $P_S(0,t|x_0)=0$.
The solution to $P_S(x,t|x_0)$ is successfully constructed using the MIs:
\begin{eqnarray}
P_S(x,t|x_0) = P_0(x,t|x_0) - P_0(x,t|-x_0),
\label{MI_solution}
\end{eqnarray}
where the second term of the right side is the contribution from an image ``anti-walker" starting at $x=-x_0$~\cite{Redner}. 
Why does it work here? 
The standard reasoning is that it solves the diffusion equation and satisfies both initial and boundary conditions.

A deeper insight can be gained by revisiting the first passage issues in terms of ``dead" walkers. Up to time $t\,(>0)$, some walkers remain in the positive domain, but others touch the boundary and are absorbed. 
Whereas we are interested in the PDF of the surviving walkers $P_S(x,t|x_0)$, the dead walkers can be introduced such that the conservation of the probability is maintained as
$P_D(x,t|x_0) \coloneqq P_0(x,t|x_0) - P_S(x,t|x_0)$, which initially leave $x_0$ at $t=0$ and then have passed the absorbing boundary ``at least once".
This provides a directly comparable form to Eq.~(\ref{MI_solution}):
\begin{eqnarray}
P_S(x,t|x_0) = P_0(x,t|x_0) - P_D(x,t|x_0).
\label{P_D}
\end{eqnarray}
Equation~\eqref{P_D} is generic and an identity in the entire domain for $x\in (-\infty, +\infty)$ regardless of the memory.
Attention should be paid to a difference in the last term on the right-hand sides between Eqs.~(\ref{MI_solution}),\,(\ref{P_D}).

\begin{figure}[b]
	\centering
	\includegraphics[width=1.00\textwidth]{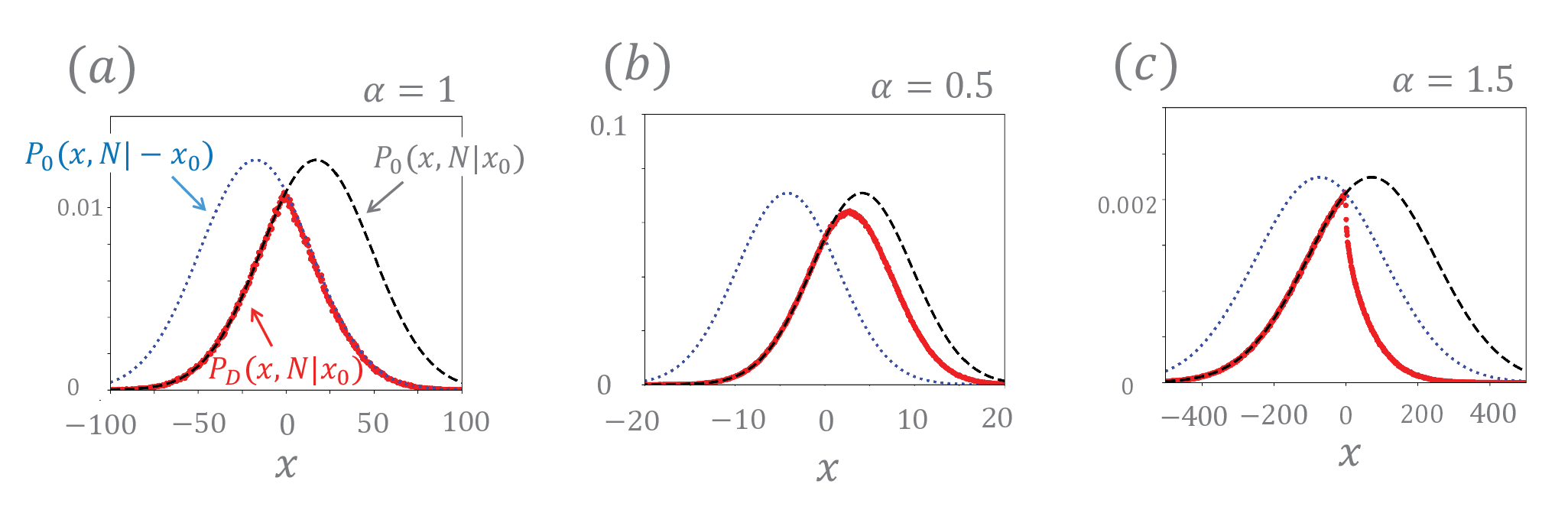}
	\caption{Plots of PDFs $P_D(x,N|x_0)$ (red), $P_0(x,N|x_0)$ (black dashed), and $P_0(x,N|-x_0)$ (blue dotted) in Eq.~\eqref{PD_P0} after $N=10^3$ steps starting from $x_0= 300^{\alpha/2}$ for (a) Markovian, (b) anti-persistent non-Markovian, and (c) persistent non-Markovian walkers.
As a model of non-Markovian walkers, we adopt the fBms with anomalous exponent $\alpha ( \neq 1)$ for mean square displacement.  $P_0(x,N|\pm x_0)$ is the Gaussian function with the mean $\pm x_0$ and the variance $2 (N \Delta t)^{\alpha}$, while $P_D(x,N|x_0)$ is obtained by numerical simulation (see Appendix~\ref{sec:3}).}
	\label{fig2}
			\vspace{0.2 cm}
\end{figure}

Figure~\ref{fig2} provides numerical demonstrations of the MI validity, where we plot $P_D(x,N|x_0)$, $P_0(x,N|x_0)$, and $P_0(x,N|-x_0)$ for the ordinary Brownian motion and the fractional Brownian motions (fBms)~~\cite{Hosking,Dieker_2004,PRE_Krug_1997,AdvPhys_Bray_2013,PRE_Amitai_2010,SoftMatter_Sokolov_2012,PCCP_Metzler_2014,JCP_Sanders_2012,PRR_Sakamoto_Sakaue_2023,Dieker_2004,PRL_Zoia_2009}.
Note that the temporal axis is discretized such that $t = N \Delta t$ with a time increment $\Delta t$ for later use, and we employ the step number $N$ in a discrete and the time $t$ in a continuous form interchangeably.
The fBm is a represented example of the nonMarkov processes, which are identified by $\left< (x_i-x_0) \right>=0$ and its covariance $\left< (x_i-x_0)(x_j-x_0) \right>=[i^\alpha +j^\alpha-|i-j|^\alpha]$ with $x_i$ being the position at $i$-th step and $\left< (\cdot) \right>$ denoting the ensemble average for free walkers.
Incidentally, for free walkers, $\alpha <1$ or $\alpha >1$ exhibits the sub- or superdiffusion, respectively, in the mean-square displacement $\left< (x_i-x_0)^2 \right>\sim i^\alpha$.
The general key relation to see Fig.~\ref{fig2} is summarized as
\begin{equation}
P_D(x,N|x_0 ) 
\begin{cases}
\neq P_0 (x,N| -x_0) & (x>0)
\\
=      P_0 (x,N| x_0) & (x<0)
\end{cases}.
\label{PD_P0}
\end{equation}
For $x<0$, any dead and free walkers leaving $x_0>0$ necessarily pass across $x=0$ and exhibit identical trajectories with the same probability.
Hence, the lower equation~(\ref{PD_P0}) is an identity holding both for the Markov and nonMarkov processes.
On the other hand, $x>0$ manifests the MI problems.
Only for the Markovian process, the equality $P_D(x,N|x_0 ) = P_0 (x,N|-x_0)$ holds.
Therein, by plugging this Markovian relation into Eq.~(\ref{P_D}), we arrive at the MI solution (Eq.~(\ref{MI_solution})).
Indeed, as in Fig.~\ref{fig2}\,(a) for the Markov processes, $P_D(x,N|x_0)$ collapses onto $P_0(x,N|x_0)$ for $x<0$ and $P_0(x,N|-x_0)$ for $x>0$, respectively.
However, Fig.~\ref{fig2}\,(b),\,(c) for the nonMarkov processes created by the fBms are not the case, i.e., $P_D(x,N|x_0 ) \neq P_0 (x,N|-x_0)$ for $x>0$, demonstrating numerical evidence of Eq.~(\ref{PD_P0}).

It is known $P_S(x,N|x_0)$ exhibits a singular behavior near the absorbing boundary with a nontrivial exponent $\delta$ in $P_S(x,N|x_0) \sim x^\delta$~\cite{PRL_Zoia_2009,PRR_Sakamoto_Sakaue_2023}. Understanding the behavior of $P_D(x,N|x_0)$ would provide valuable insights for the singularity, as these two quantities $P_S(x,t|x_0)$, $P_D(x,t|x_0)$ are linked via Eq.~(\ref{P_D}).

\section{Probability ratio between dead and reflected paths}
\label{FT_formalism_sec}

Section~\ref{FT_formalism_sec} focuses attention on mathematical MI structures of general Gaussian processes that include nonMarkov as well as Markov processes.

We now employ the discrete form to clearly show matrix operations, while a continuous form provides a direct form of the GLE.
The position is updated in discrete time according to
\begin{eqnarray}
\Delta x_i = \xi_i,
\label{EM_vel}
\end{eqnarray}
where $\Delta x_i\coloneqq x_i-x_{i-1}$, and $\xi_i$ represents noise at $i$-th step with the mean $\left< \xi_i \right>=0$ and its covariance $C_{ij} = \langle \xi_i \xi_j \rangle $ being a symmetric matrix $C = (  C_{ij}) \in \mathbb{R}^{N \times N}$.
A stochastic position path $\{ x_{i} \}_0^{N}=(x_0, x_1, \cdots, x_N)$ with $x_i = x_0 + \sum_{k=1}^i \Delta x_k$ is generated by iteratively substituting a noise from the sequence $\{ \xi_i \}_1^N=\{ \xi_1, \xi_2, \cdots, \xi_N \}$ into Eq.~(\ref{EM_vel}) given the initial position $x_0$.
Bear in mind that Eq.~(\ref{EM_vel}) intrinsically corresponds to the relation in the velocity space with $v_i \coloneqq \Delta x_i/\Delta t$.
The path probability of $\{ x_{i} \}_0^{N}$~\cite{PhysRev_Onsager_Machlup_1953,JStatMech_Ohkuma_Ohta_2007} is found by transforming the variables from $\{ \xi_i \}_1^N$ into $\{ \Delta x_i \}_1^N$ with Eq.~(\ref{EM_vel}) along the Onsager-Machlup idea~\cite{PhysRev_Onsager_Machlup_1953,JStatMech_Ohkuma_Ohta_2007} (cf. Appendix~\ref{sec:0}), so that the probability of the noise sequence $P_0^{(\xi)}[\{ \xi_i \}_1^{N}]\sim  \exp{[ -(1/2)\sum_{i=1}^N\sum_{j=1}^N C_{ij}^{-1}\xi_i\xi_j  ]}$ then turns into
\begin{eqnarray}
P_0[\{ x_{i} \}_0^{N}]
\sim
 \exp{[ -I[\{ \Delta x_{i} \}_1^{N}] ]}
\label{P}.
\end{eqnarray}
The Jacobian $J=|\partial \xi_j/\partial \Delta x_i|$ drops in Eq.~(\ref{P}) because it is irrelevant to the following formulations.
In addition, the $N$-step process is denoted by a ket as a column vector $\ket {\Delta x} = (\Delta x_1, \Delta x_2, \cdots, \Delta x_N)^\top \in \mathbb{R}^N$, the inverse matrix is defined as $\Gamma = C^{-1}$ to satisfy $\sum_{i=1}^N C_{ij}\Gamma_{jk}=\delta_{ik}$, and we introduce action $I[\{ \Delta x_{i} \}_1^{N}] =   \bra {\Delta x} \Gamma \ket { \Delta x} /2$~\cite{AdvPhys_Roldan_2023}.
For example, in the simple random walk corresponding to the Brownian particle, the action is identified as $I[\{ \Delta x_{i} \}_1^{N}] ]=(1/4D)\sum_{i=1}^N \Delta x_i^2$, where the diagonal matrices $C_{ij}=2D\delta_{ij}$ and $\Gamma_{ij}=C_{ij}^{-1}=(1/2D)\delta_{ij}$ with $D$ being a diffusion constant. 
This instructive example demonstrates a well-known linear proportion of the mean-square displacement $\left< (x_i-x_0)^2 \right>=2Di$.

We consider the decomposition of probability (see Eq.~(\ref{P_D})) at the path-probability level;
\begin{eqnarray}
P_S[\{ x_{i} \}_0^{N}]=P_0[\{ x_{i} \}_0^{N}]-P_D[\{ x_{i} \}_0^{N}],
\label{P_D_path}
\end{eqnarray}
where $P_S[\{ x_{i} \}_0^{N}]=\prod_{i'=1}^N \Theta^+(x_{i'})P_0[\{ x_{i} \}_0^{N}]$ and $P_D[\{ x_{i} \}_0^{N}]=\left(1-\prod_{i'=1}^N \Theta^+(x_{i'})\right)P_0[\{ x_{i} \}_0^{N}]$. Note that $\Theta^+(x)$ is a step function taking $\Theta^+(x)=0$ for $x\leq 0$, otherwise $\Theta^+(x)=1$.
Formally, Eq.~(\ref{P_D_path}) is verified by inserting an identity $\prod_{i'=1}^N \left[ \Theta^+(x_{i'})+\Theta^+(-x_{i'}) \right]=1$ into $P_0[\{ x_{i} \}_0^{N}]$ and expand it.

We now analyze the path probability of the dead walkers.
Let $\{ x_{i} \}_{0,n}^{N} \coloneqq (\{ x_{i} \}_0^{n}, \{ x_{i} \}_{n+1}^{N}) $ be a ``dead path" that first touches the boundary at time $n\, (\le N)$.
This separation also decomposes the velocity-sequence vector $\ket {\Delta x} \coloneqq (\ket {\Delta x'}, \ket {\Delta x''} )$
where $\ket {\Delta x'} = (\Delta x_1, \Delta x_2, \cdots, \Delta x_n)^\top \in \mathbb{R}^n$ and  $\ket {\Delta x''} = (\Delta x_{n+1}, \Delta x_{n+2}, \cdots, \Delta x_N)^\top \in \mathbb{R}^m$ represent the initial $n$ steps up to the first arrival and the remaining $m\, (\coloneqq N-n)$ steps, respectively.
This restrictive description allows us not to explicitly deal with the dead-path filter: $\left(1-\prod_{i'=1}^N \Theta^+(x_{i'})\right)\,(=P_D[\{ x_{i} \}_0^{N}]/P_0[\{ x_{i} \}_0^{N}])$, so that the investigations of the MIs with the dead paths is reduced to analyses with block matrices:
\begin{eqnarray}
C = 
\begin{pmatrix}
  C^{\mathrm{I}}& B \\
  B^\top & C^{\mathrm{II}}
\end{pmatrix},
\quad
\Gamma = 
\begin{pmatrix}
  \Gamma^{\mathrm{I}}& \Omega \\
  \Omega^\top & \Gamma^{\mathrm{II}}
\end{pmatrix},
\label{block_M}
\end{eqnarray}
where $C ^{\mathrm{I}} \in \mathbb{R}^{n \times n}$, $C ^{\mathrm{II}}\in \mathbb{R}^{m \times m}$, $B  \in \mathbb{R}^{n \times m}$ and $B^\top  \in \mathbb{R}^{m \times n}$ is the transpose of $B$. The same decomposition applies to $\Gamma$ (cf. block matrix operations~\cite{Kuttler}).
The action is now written as $I[\{ \Delta x_{i} \}_{1,n}^{N}] = I^{\mathrm{I}}[\{ \Delta x_{i} \}_1^{n}] + I^{\mathrm{II}}[\{ \Delta x_{i} \}_{n+1}^{N}]+  \Lambda [\{ \Delta x_{i} \}_{1,n}^N ]$ with
\begin{eqnarray}
&&I^{\mathrm{I}} [ \{ \Delta x_{i} \}_{1}^n] = \frac{1}{2} \bra{\Delta x'} \Gamma^{\mathrm{I}} \ket{\Delta x'} \nonumber \\
 &&I^{\mathrm{II}} [ \{ \Delta x_{i} \}_{n+1}^N] =\frac{1}{2} \bra{\Delta x''} \Gamma^{\mathrm{II}} \ket{\Delta x''}    \nonumber \\
 &&\Lambda [ \{ \Delta x_{i} \}_{1,n}^N] =    \bra{\Delta x''} \Omega^\top \ket{\Delta x'}=\bra{\Delta x'} \Omega \ket{\Delta x''}.
\label{H_2}
\end{eqnarray}

To discuss a connection with the MIs, we define a path 
\begin{eqnarray}
 \{ \overline{x}_{i} \}_{0,n}^{N}  \coloneqq (\{ -x_{i} \}_0^{n}, \{ x_{i} \}_{n+1}^{N}),
\label{x_dead}
\end{eqnarray}
hereafter called ''reflected dead path" or simply ''reflected path", which is obtained from the original dead path by spatially reflecting the first passage path (see Fig.~\ref{fig1}). 
Recall that, throughout the article, the reflection point is fixed at the absorbing boundary ($x=0$).
Since only $\Lambda$ in Eq.~\eqref{H_2} is odd under the spatial reflection while other terms are even, one finds the probability ratio between the original and its reflected paths
\begin{eqnarray}
\frac{P_D [\{ x_{i} \}_{0,n}^{N}]}{P_{\overline{D}} [\{ \overline{x}_{i} \}_{0,n}^{N}]} = \exp{[-2 \Lambda [\{  \Delta x_i\}_{1,n}^N}]].
\label{path_ratio}
\end{eqnarray}
For Markovian processes, $C$ and $\Gamma$ are diagonal matrices; hence, all the off-diagonal block matrices $B$ and $\Omega$ are null. Since Eq.~\eqref{path_ratio} then states that the original and reflected paths are equally probable, i.e., $\Lambda [\{  \Delta x_i\}_{1,n}^N]=0$, summing over all the pertinent paths, we obtain
\begin{eqnarray}
P_D(x,N|x_0 ) = P_{\overline{D}}(x,N|\overline{x}_0), \quad (\mathrm{Markovian})
\label{Markov_D_Dbar}
\end{eqnarray} where $P_{\overline{D}}(x,N|\overline{x}_0)$ is the PDF of the reflected dead walkers. In such reflected paths, the walker starts at $x=\overline{x}_0 ( = -x_0)$ and crosses the boundary at $x=0$ to reach the final position $x\,(>0)$ after $N$ steps. Since such paths with $\overline{x}_0<0$ and $x>0$ necessarily cross the boundary, these paths are unrestricted; hence, 
\begin{eqnarray}
P_{\overline{D}}(x,N|\overline{x}_0) = P_0 (x,N|\overline{x}_0),  \qquad (x_0 >0 , \ x>0)
\label{P_barD_P_0-}
\end{eqnarray}
which is generally valid regardless of Markovity.
Combining Eq.~(\ref{Markov_D_Dbar}) with Eq.~(\ref{P_barD_P_0-}) follows the MI construction $P_D(x,N|x_0 ) =P_0 (x,N|\overline{x}_0)$ for the Markov processes, but not for the nonMarkov processes because of the MI violation factor $\Lambda [\{  \Delta x_i\}_{1,n}^N] \neq 0$.

\section{Energetic interpretation}
\label{TD_sec}

We now consider that the system is in contact with a thermal bath and assume the FDR such that a class of stochastic processes, like the fBms, can be described using GLE.
This allows us to make the main claim that the quantity $\Lambda [ \{ \Delta x_i\}_{1,n}^N] $ appearing in Eq.~(\ref{path_ratio}) is related to heat produced by the memory force in the nonMarkov processes.
Throughout the paper, we adopt a unit system, where thermal energy is set to unity. 

A hint to the heat is an apparent similarity of Eq.~\eqref{path_ratio} to the local detailed balance, a key relation in the fluctuation theorem~\cite{Maes_2021,PRE_Crooks_2000,RPP_Seifert_2012,PRX_Jarzynski_2017,AdvPhys_Roldan_2023,JStatMech_Chernyak_Chertkov_Jarzynski_2006,JStatMech_GarciaGarcia_2012}.
Recall that the conventional fluctuation theorem combines the temporal reverse path with the original, whereas in the present context a pair of paths is constructed by the spatial reflection (cf. classifications by entropic functionals in Appendix~\ref{Class_FT}).

\subsection{$\Lambda=\varDelta' Q$}

\begin{figure}[b]
	\centering
	\includegraphics[width=1.00\textwidth]{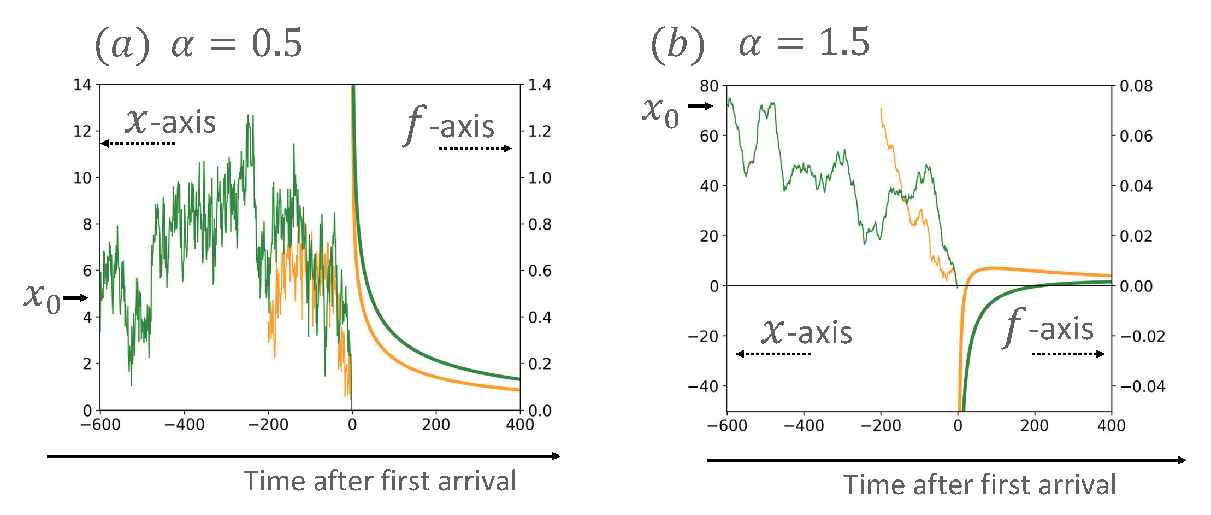}
	\caption{Examples of first passage paths and resultant memory forces acting on the walker in the subsequent process for (a) subdiffusive fBm ($\alpha = 0.5$) and (b) superdiffusive fBm ($\alpha=1.5$). The time axis is shifted to set the moment of the first passage to be the origin.}
	\label{fig3}
			 \vspace{0.2 cm}
\end{figure}

Let us fix a specific first passage path $\{ x_{i} \}_0^{n}$ and inquire into its consequences on the subsequent dynamics.
To this end, we rewrite the action for the discrete form as
\begin{eqnarray}
I[ \{ \Delta x_{i} \}_{1,n}^N] = I_{\mathrm{eff}} [ \{ \Delta x_{i} \}_{1,n}^N] + I_{(0)}^{\mathrm{I}} [ \{ \Delta x_{i} \}_{1}^n],
\end{eqnarray}
where the effective action is introduced as
\begin{eqnarray}
I_{\mathrm{eff}} [ \{ \Delta x_{i} \}_{1,n}^N] &=& \frac{1}{2} \biggl[\bra{\Delta x''} - \bra{f} (\Gamma^{\mathrm{II}})^{-1}\biggr] \Gamma^{\mathrm{II}} \biggl[ \ket{\Delta x''} -(\Gamma^{\mathrm{II}})^{-1}\ket{f}  \biggr] 
\label{H_eff}
%H^{\mathrm{I}} (\{ \Delta y_i\}) &=&  \frac{1}{2} \bra{\Delta y } (D^{\mathrm{I}})^{-1} \ket{\Delta y} 
\end{eqnarray}
with $ I_{(0)}^{\mathrm{I}} [ \{ \Delta x_{i} \}_{1}^n] =  \bra{\Delta x'} (C^{\mathrm{I}})^{-1} \ket{\Delta x'}/2$ and 
\begin{eqnarray}
\ket{f[ \{ \Delta x_{i} \}_{1,n}^N] } \coloneqq - \Omega^\top \ket{\Delta x'} \in \mathbb{R}^{m}.
\label{memory_f}
\end{eqnarray}
%% (refer to Appendix~\ref{sec:0} with Eqs.~(\ref{GLE_eff})-(\ref{Lambda_def_Lang}) for the continuous GLE form).
Note that the redundant arguments drop like $\ket{f[ \{ \Delta x_{i} \}_{1,n}^N] } \rightarrow \ket{f}$.
This form results from completing the square with respect to $\ket{\Delta x''}$, where the transformation introduces a ``memory force" $\ket{f}$ depending on the specific first passage path $\ket{\Delta x'}$ (see Eq.~(\ref{memory_f})).
Examples of such a force are shown in Fig.~\ref{fig3} for cases of sub-diffusive and super-diffusive fBms.
The conditional  path probability given the first passage path $\{ x_{i} \}_0^{n}$ is then identified as
\begin{eqnarray}
P_D[\{ x_{i} \}_n^{N}| \{ x_{i}\}_0^n]
\sim
 \exp{[ -I_{\mathrm{eff}} [ \{ \Delta x_{i} \}_{1,n}^N] ]}.
\label{P_II}
\end{eqnarray}
It is worth noting that the matrix $\Gamma^{\mathrm{II}}_{ij}$ in the quadratic form of Eq.~(\ref{H_eff}) is related to the noise statistics $\{ x_i \}_{n+1}^N$ through the FDR of second kind.
This implies the following relation: $\ket{\Delta x''} = (\Gamma^{\mathrm{II}})^{-1} \ket{f} + \ket{\xi^{\mathrm{II}}} $, where $\ket{\xi^{\mathrm{II}}} \in \mathbb{R}^{m}$ is the noise vectors with $\langle \xi_i^{\mathrm{II}} \rangle =0$, and $\langle \xi_i^{\mathrm{II}} \xi_j^{\mathrm{II}} \rangle = (\Gamma^{\mathrm{II}})^{-1}_{ij}$.
By recalling $m=N-n$, its $m$-th component, $\Delta x''_m = \sum_{j=1}^{m}(\Gamma^{\mathrm{II}})^{-1}_{mj} f_j + \xi_m^{\mathrm{II}}$ describes the time evolution from $x_{N-1}$ to $x_{N}$ under a given history $\{ x_{i} \}_0^{N-1}$, which is analogous to the algorithm of Hosking method widely used to generate nonMarkovian Gaussian trajectories in numerical simulations~\cite{Hosking,Dieker_2004}.
 Operating $\Gamma^{\mathrm{II}}$ from the left to the preceding vector equation, we obtain $\Gamma^{\mathrm{II}}\ket{\Delta x''} =  \ket{f} + \ket{\zeta}$ with  $\ket{\zeta} = \Gamma^{\mathrm{II}} \ket{\xi^{\mathrm{II}}} \in \mathbb{R}^{m}$, hence, $ \langle \zeta_i \rangle =0$ and the FDR $\langle \zeta_i \zeta_j \rangle = \Gamma^{\mathrm{II}}_{ij}$.
Its $m$-th component takes the discrete form of the GLE with the memory force $f_m\coloneqq (\ket{f})_m$:
\begin{eqnarray}
\sum_{j=1}^m \Gamma^{\mathrm{II}}_{mj} \  \Delta x''_j &=& f_m + \zeta_m,
\label{GL_1}
\end{eqnarray}
which represents the ``force balance" at time $N = n+m$ given the past history $\{ x_{i} \}_0^{N-1}$. One can repeat the above argument by replacing $m$ with $m' = N' -n \ (<m)$ along with the corresponding redefinitions of all the matrices and vectors ($\Gamma ^{\mathrm{II}}\in \mathbb{R}^{m' \times m'}$, $\Omega^\top  \in \mathbb{R}^{m' \times n}$, $\ket{f} \in \mathbb{R}^{m'}$ etc.), whereby construct the force balance equation at time $N' = n+m'$.
Following the stochastic energetic definition of the heat~\cite{Sekimoto_book}, the quantity
\begin{eqnarray}
\Delta' Q_{m'} &\coloneqq& \left(-\sum_{j=1}^{m'} \Gamma^{\mathrm{II}}_{m'j} \  \Delta x''_j  + \zeta_{m'} \right) \Delta x''_{m'} = -f_{m'} \Delta x''_{m'}
\label{heat_rate_discrete}
\end{eqnarray}
is interpreted as the absorbed heat in the step $x_{N'-1}$ to $x_{N'}$. Therefore, using Eq.~(\ref{memory_f}), we identify $\Lambda [ \{ \Delta x_{i} \}_{0,n}^N]
= \bra{\Delta x''}\Omega^\top \ket{\Delta x'}
=-  \braket{\Delta x'' |  f} =  \sum_{m'=1}^m  \Delta' Q_{m'}$ as the total heat to the system in the post-first-passage path $\{ x_{i} \}_n^{N}$.

Here, key quantities expressed by the GLE in the continuous form are briefly shown below (see also Appendix A), to contrast them with the discrete form.
In doing so, we recall the following notational correspondences between the block and the original matrices (see $\Gamma$ in Eq.~(\ref{block_M}));
\begin{eqnarray}
\Gamma^{\mathrm{II}}_{ij} &=& \Gamma_{n+i,n+j} \quad (1 \leq i,j \leq m) 
\nonumber \\
(\Omega^\top)_{ij} &=& \Gamma_{n+i,j} \quad (1 \leq i \leq m, \,1 \leq j \leq n) 
\nonumber \\
\Delta x_j' &=& \Delta x_j \quad (1 \leq j \leq n) 
\nonumber \\
\Delta x_j'' &=& \Delta x_{n+j} \quad (1 \leq j \leq m).
\label{blcok_original_matrix}
\end{eqnarray}
Then, the memory force at time $t=N\Delta t$ is given by the continuum limit of the $m$-th component of Eq.~(\ref{memory_f}):
\begin{eqnarray}
f_t \ \coloneqq -\int_0^\tau ds\, \Gamma_{ts}\dot{x}_s,
\label{memory_f_continuum}
\end{eqnarray}
where $\tau \coloneqq n \Delta t$ is the first passage time.
Parallelly, Eq.~(\ref{GL_1}) is replaced by the GLE:
\begin{eqnarray}
\int_\tau^t ds\,\Gamma_{ts}\dot{x}_s &=& f_t+\zeta_t.
\label{GLE_eff}
\end{eqnarray}
As a correspondence to Eq.~(\ref{heat_rate_discrete}), a heat rate for the GLE is
\begin{eqnarray}
\frac{d'Q}{dt}
&=&
\left(
-\int_\tau^t ds\,\Gamma_{ts}\dot{x}_s +\zeta_t
\right) \dot{x}_t
=
-f_t \dot{x}_t,
\label{GLE_heat}
\end{eqnarray}
where the first equality is the definition of the heat rate, and the last equation is obtained by substituting Eq.~(\ref{GLE_eff}).
A remarkable point is that the FDR $\left< \zeta_s\zeta_{s'} \right>=\Gamma_{ss'}$ holds (see the covariance and its inverse (Green function) for $s,s'\in [\tau,t]$ in Eq.~(\ref{PD_completeSq})).
The correlation between the first-passage and the post-first-passage paths given by Eq.~(\ref{H_2}) is redefined as
\begin{eqnarray}
\Lambda[\{ \dot{x}_{t'} \}_{0,\tau}^t]
&\coloneqq&
\int_\tau^t ds\,\int_0^\tau ds'\,
\Gamma_{ss'}\dot{x}_s\dot{x}_{s'},
\label{Lambda_def_Lang}
\end{eqnarray}
which is shown to be related to the heat using Eqs.~(\ref{memory_f_continuum}),\,(\ref{GLE_heat}).

\subsection{Analogous Relations to Fluctuation Theorems}

Now we discuss analogous relations to fluctuation theorems by going back to the discrete notation. 
Noting $\Lambda[\{\cdot\}_{1,n}^N]$ in Eq.~\eqref{path_ratio} is odd $\Lambda[\{ \Delta x_i\}_{1,n}^N] = -\Lambda[\{ \Delta \overline{x}_i\}_{1,n}^N] $ under the reflection (see Eq.~(\ref{x_dead})), one then pursues a standard line of argument by performing integral over dead paths with no restriction on the end-point $x_N =x\in (-\infty, \infty)$ to obtain 
\begin{eqnarray}
 \frac{P_D(\Lambda,N|x_0)}{ P_{\overline{D}}(-\Lambda,N|\overline{x}_0)}=    \exp{(-2 \Lambda)}.
 \label{ft_Lambda}
\end{eqnarray}
It follows the integral fluctuation theorem $\langle e^{2 \Lambda} \rangle_D = 1$ for $\Lambda$ with the ensuing inequality $\langle \Lambda \rangle_D \le 0$ reminiscent of the second law (refer to Appendix~\ref{sec:1} for the details with the averaging $\langle (\cdot) \rangle_D$ etc.).

The original or the reflected dead probability that takes $\Lambda$ or $-\Lambda$, respectively, but terminates at the same position $x$ at $N$ steps satisfies
\begin{eqnarray}
\frac{P_D(x,\Lambda, N|x_0) }{P_{\overline{D}}(x, -\Lambda, N| \overline{x}_0)}
 =\exp{(-2 \Lambda)}.
\label{DF_2}
\end{eqnarray}
Integrating Eq.~\eqref{DF_2} over $\Lambda$ then finds
\begin{eqnarray}
\frac{ P_D(x,N|x_0)}{ P_{{\overline{D}}}(x,N|\overline{x}_0)} =\langle  \exp{(2 \Lambda)} \rangle_{\overline{D},x}& =&\langle  \exp{(2 \Lambda)} \rangle_{D,x}^{-1}.
\label{vi_IM}
\end{eqnarray}
Introducing the Shannon entropy for the dead or the reflected dead walkers as $S_D(x,N|x_0)\coloneqq -\ln{P_D(x,N|x_0)}$ or $S_{\overline{D}}(x,N|\overline{x}_0)\coloneqq -\ln{P_{\overline{D}}(x,N|\overline{x}_0)}$, respectively, we can replace the relation~(\ref{vi_IM}) into 
\begin{eqnarray}
S_D(x,N|x_0)
-
S_{\overline{D}}(x,N|\overline{x}_0)
&=&
\ln{\langle  \exp{(2 \Lambda)} \rangle_{D,x}}
=
-\ln{\langle  \exp{(2 \Lambda)} \rangle_{\overline{D},x}}.
\label{IFT_Lambda}
\end{eqnarray}
Applying Jensen's inequality, we obtain
\begin{eqnarray}
S_D(x,N|x_0)
-
S_{\overline{D}}(x,N|\overline{x}_0)
&\geq&
2 \langle \Lambda \rangle_{D,x}
\nonumber \\
S_{\overline{D}}(x,N|\overline{x}_0)
-
S_D(x,N|x_0)
&\geq&
2 \langle \Lambda \rangle_{{\overline{D}},x}
\label{Lambda_Clausius}
\end{eqnarray}
The above relations bound the excess entropy flowed by the space reflection with the heat, and may be interpreted as the Clausius inequality in the MI violation.

Let us now recall the story of the MI violations with Fig.~\ref{fig2}.
By imposing then the endpoint to be in the physical domain ($x \ge 0$) in Eqs.~(\ref{vi_IM}),\,(\ref{IFT_Lambda}),\,(\ref{Lambda_Clausius}), the use of Eq.~\eqref{P_barD_P_0-} allows us to replace $P_{\overline{D}}(x,N|\overline{x}_0)$ with $P_{0}(x,N|\overline{x}_0)$.
This provides explicit representations to quantify or bound the departure from the MI result in non-Markovian processes in terms of the heat (see Sec.~\ref{subsec_Memory_f}). In a similar way, one can obtain the expression of the survival probability in terms of the exponential average of the heat (see Appendix~\ref{sec:2}).

\begin{figure}[b]
	\centering
	\includegraphics[width=1.0\textwidth]{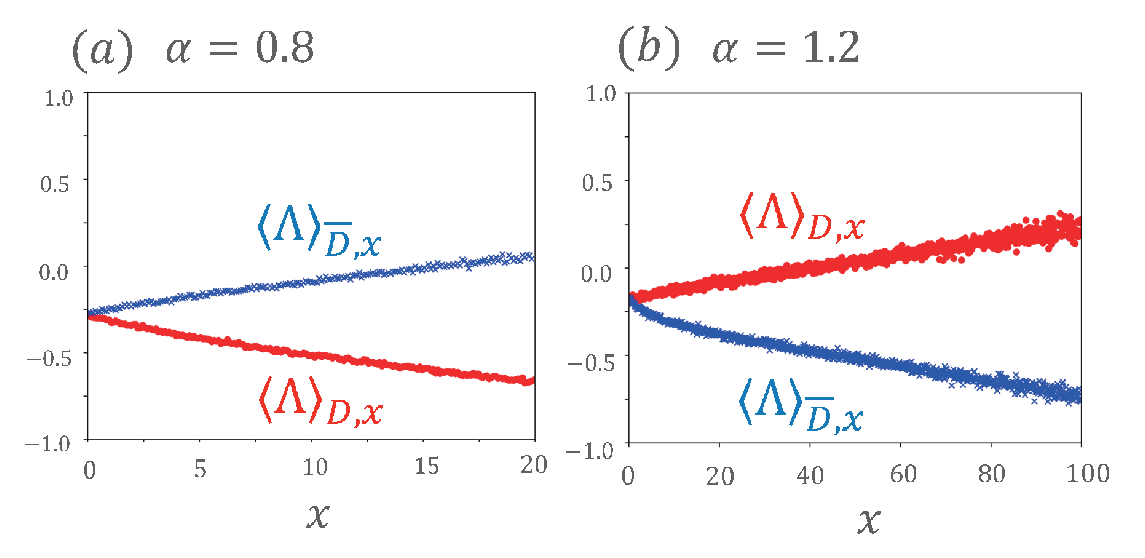}
	\caption{Average heat flow in the post-first-passage process. (a) fBm with $\alpha=0.8$ and (b) fBm with $\alpha=1.2$. The starting position is $x_0=300^{\alpha/2}$.
	Average absorbed heat $  \langle \Lambda \rangle_{D,x}$ along the dead paths (red)  and $ \langle \Lambda \rangle_{{\overline D},x}$ along the reflected paths (blue) are shown as functions of the endpoint $x\, (\ge 0)$ after $N=10^3$ steps.  
	Note that $\langle \Lambda \rangle_{D,0}=  \langle \Lambda \rangle_{{\overline{D}},0}$ holds because the path probabilities have a spatial symmetricity $P_D[\{ x_i \}_1^N | x_N=0, x_0]=P_{\overline{D}}[\{ -x_i \}_1^N | -x_N=0, \overline{x}_0]$ by reflecting the entire path as $x_i \rightarrow -x_i$ for $i \in [0,N]$.
}
	\label{fig4}
			\vspace{0.2 cm}
\end{figure}

\section{Discussions}
\label{Discussion_sec}
Compared to Markovian systems, the behaviors of non-Markovian systems are more involved in many respects. A distinct feature of the latter can be seen in their conditional behaviors based on past history. With such information encoded, the PDF $P_D(x,N|x_0)$ of the dead walkers is non-trivial and difficult to analytically specify, whereas the unconditional PDF $P_0(x,N|x_0)$ is not.
Indeed, the energetics of the unconditional non-Markovian process in the absence of the potential landscape is as trivial as the corresponding Markovian process, i.e., in both cases, the viscous (or viscoelastic) resistance force and the noise balance to zero in the overdamped limit, hence, no net energy flow between the system and the
environment. For conditional non-Markovian processes, however, we have just seen that the process is biased due to the memory, putting the resulting energetics due to such a biasing force on the stage. Here, the relevant quantity is $\Lambda$, the heat flow to the system in the post-first-passage process.

\subsection{Memory force quantified by $\Lambda$}
\label{subsec_Memory_f}

Let us examine the statistical properties of $\Lambda$. 
Figure~\ref{fig4} plots the average heat in the post-first-passage process as a function of its endpoint $x=x_N$ for subdiffusive and superdiffusive fBms. Recall that $\langle \Lambda \rangle_{D, x}$ is the average heat absorption during the post-first-passage process in the dead paths conditioned on a nonnegative endpoint $x=x_N \ge 0$.
Of also interest is the corresponding quantity $\langle \Lambda \rangle_{{\overline D}, x}$ for the reflected paths that start at ${\overline x}_0 < 0$ conditioned on the same end $x=x_N \ge 0$ as the dead one.
As we have stated, Eq.~(\ref{Lambda_Clausius}) for the use in $x \geq 0$ is written as
\begin{eqnarray}
\langle \Lambda \rangle_{{\overline{D}},x} \le \frac{1}{2} \ln{\left( \frac{P_D(x,N|x_0)}{P_0(x,N|\overline{x}_0)} \right)} \le - \langle \Lambda \rangle_{D,x},
\label{MI_bound}
\end{eqnarray}
which provides bounds on the violation of the MI result in terms of the heat.
We quite generally observe that the heat emission on average into the environment $-\langle \Lambda \rangle_{D, x} > 0$. It is particularly true for anti-persistent processes, as might be naturally expected from $P_D(x,N|x_0) \ge P_0(x,N|\overline{x}_0)$ (Fig.~\ref{fig2}(b)). Such a naive intuition, however, does not apply straightforwardly to a persistent process, for which $P_D(x,N|x_0) \le P_0(x,N|\overline{x}_0)$ (Fig.~\ref{fig2}(c)).
As shown in Fig.~\ref{fig4}, 
while the amount of heat emission increases with $x$ for $\alpha < 1$, it decreases with $x$ for $\alpha > 1$ and reverse sign for large $x$.
The latter result for $\alpha>1$ indicates that, after the dead walker leaving $x_0 > 0$ touches the boundary at $x=0$, reaching far in the positive domain requires heat absorption. We note that the positive $\langle \Lambda \rangle_{D, x}$ for large $x$ is compatible with the inequality~\eqref{MI_bound}, which ensures the positivity for the sum $-(\langle \Lambda \rangle_{D, x} + \langle \Lambda \rangle_{{\overline{D}}, x})$, but not for individual terms.

As observed in a tagged-monomer system projected from a polymer chain~\cite{Panja_2010,PRE_Sakaue_2013,Maes_2013,PRE_Saito_Sakaue_2015,Vandebroek_2017,PRE_Saito_2022,PRL_Gupta_2013}, non-Markovian processes usually imply the presence of hidden degrees of freedom, whose dynamics are slow enough in comparison to the observed degrees of freedom, i.e., the system.
The system exchanges energy with such a slowly fluctuating environment in the form of either heat or work. 
To make sense of such an energetic picture, the present results can be summarized into the following formulation, where the continuous form is employed for convenience.
During the initial first passage process $\{ x_{t'}\}_{0}^\tau$, the system interacts with the environment and builds correlation with it, creating a potential $U(x,t |  \{ x_{t'} \}_{0,\tau}^t ) = - f_t x$ for the post first passage process $\{ x_{t'}\}_{\tau}^t$.
Since integration by parts of $\Lambda[\{ \dot{x}_{t'} \}_{0,\tau}^t]=-\int_\tau^tds\,f_s\dot{x}_s$ finds $U(x,t | \{ x_{t'} \}_{0,\tau}^t )=\Lambda[\{ \dot{x}_{t'} \}_{0,\tau}^t] -\int_\tau^tds\,(\partial_s f_s)x_s$,
the energy balance during the stochastic time evolution over time period $dt$~\cite{Sekimoto_book} is written as
\begin{eqnarray}
dU_t = -f_t dx_t - (\partial_t f_t) x_t dt,
\label{dU_nonMarkov}
\end{eqnarray}
where compact notations are adopted like $U(x,t|\{ x_{t'} \}_{0,\tau}^t )\rightarrow U_t$ by dropping some arguments.
Our definition of absorbed heat $d'Q_t =- f_t dx_t$ indicates $d'W_t \coloneqq  - (\partial_t f_t) x_t dt$ to be identified as the work done to the system.
In anti-persistent process, $f_t$ is positive (see Fig.~\ref{fig3}\,(a)), thus, upon the first passage to $x=0$, the walker tends to be pushed back into positive domain as reflected in $P_D(x,N|x_0) > P_0(x,N|\overline{x}_0)$  (Fig.~\ref{fig2}(b)). Since the push-back motion corresponds to sliding down the potential, it accompanies the heat emission, but the temporally decaying nature of the force $f_t$ indicates lifting up the potential (Eq.~(\ref{dU_nonMarkov})) in the positive domain, hence the work done to the system. This explains the observation $\langle \Lambda \rangle_{D,x=0} < 0$ even without net change in the potential value for such paths $x_n=0 \rightarrow x_N=0$ (see also Appendix~\ref{sec:3}).

In contrast, $f_t$ is initially negative in a persistent process, whose magnitude decays with time (Fig.~\ref{fig3}\,(b)). One might expect that the walker needs to absorb heat in climbing the potential to enter the positive domain. However, the inspection reveals that the walker typically slides down the potential at first toward the negative domain, providing an opportunity for the work due to potential lift to be done. Given this work, the walker needs to absorb less heat to later reach the positive domain. In addition, $f_t$ becomes positive at later times in typical cases (Fig.~\ref{fig3}), though its magnitude is small, which contributes to making the net heat absorption negative. Still, the walker needs to absorb heat to reach far into the positive domain, as exemplified in Fig.~\ref{fig4}. These observations indicate the importance of relatively minor paths accompanying positive heat absorption, cf.  Figs~\ref{figS1},\,\ref{figS3} in Appendix~\ref{sec:3}, which dominate the exponential average $\langle  \exp{(2 \Lambda)} \rangle_{D,x} > 1$
%%  $\langle  \exp{(2 \Lambda)} \rangle_{D,x} > 0$
 for $x>0$ in Eq.~\eqref{vi_IM}, hence establish the spatial profile $P_D(x,t|x_0)$ of the dead walkers, see Fig.~\ref{fig2} (c).

\subsection{$\Lambda$ derived from noise alternations}

Besides the path probability argument made so far, another view to arrive at the memory force ``$f$" is a conversion from frictional force to noise due to the temporal-origin shift in GLE, e.g., as, in continuous form, $\int_0^t ds\,\Gamma(t-s)\dot{x}_s=\zeta_t~\rightarrow~\int_\tau^t ds\,\Gamma(t-s)\dot{x}_s=\zeta'_t$ with mean zero $\left<\zeta_t\right>=0$, the variance $\left<\zeta_t\zeta_s\right>=\Gamma(t-s)$, and the path-encoded noise $\zeta'_t\coloneqq \zeta_t-\int_0^\tau ds\,\Gamma(t-s)\dot{x}_s$ with $0<\tau<t$.
The noise difference accounts for the heat identified by $\Lambda = \int_{\tau}^t dt'\,(\zeta_{t'}-\zeta'_{t'})\dot{x}_{t'}$.

\subsection{Inhomogeneous processes}

The GLEs and its discrete variants towards the processes driven by nonstationary noises can be made.
The noise correlations $C_{t,t'}=\left< \xi_t\xi_{t'} \right>$ of the ordinary Brownian motion and also the fBms retain translational temporal symmetry, but our analyses in Secs.~\ref{FT_formalism_sec},\,\ref{TD_sec} do not assume it.
Hence, the arguments in Secs.~\ref{FT_formalism_sec},\,\ref{TD_sec} are readily found to apply to inhomogeneous Gaussian processes with nonstationary increments.
Specifically, this extension includes not only the scheduled, but also the stochastically varying correlations like telegraphic multifractional Brownian motions~\cite{PRL_Blcerek_2025}, where the $\alpha$ or the Hurst exponent to characterize the mean-square displacement is not temporally fixed, but stochastically varies according to telegraph processes or their variants.

\begin{figure}[b]
	\centering
	\includegraphics[width=1.0\textwidth]{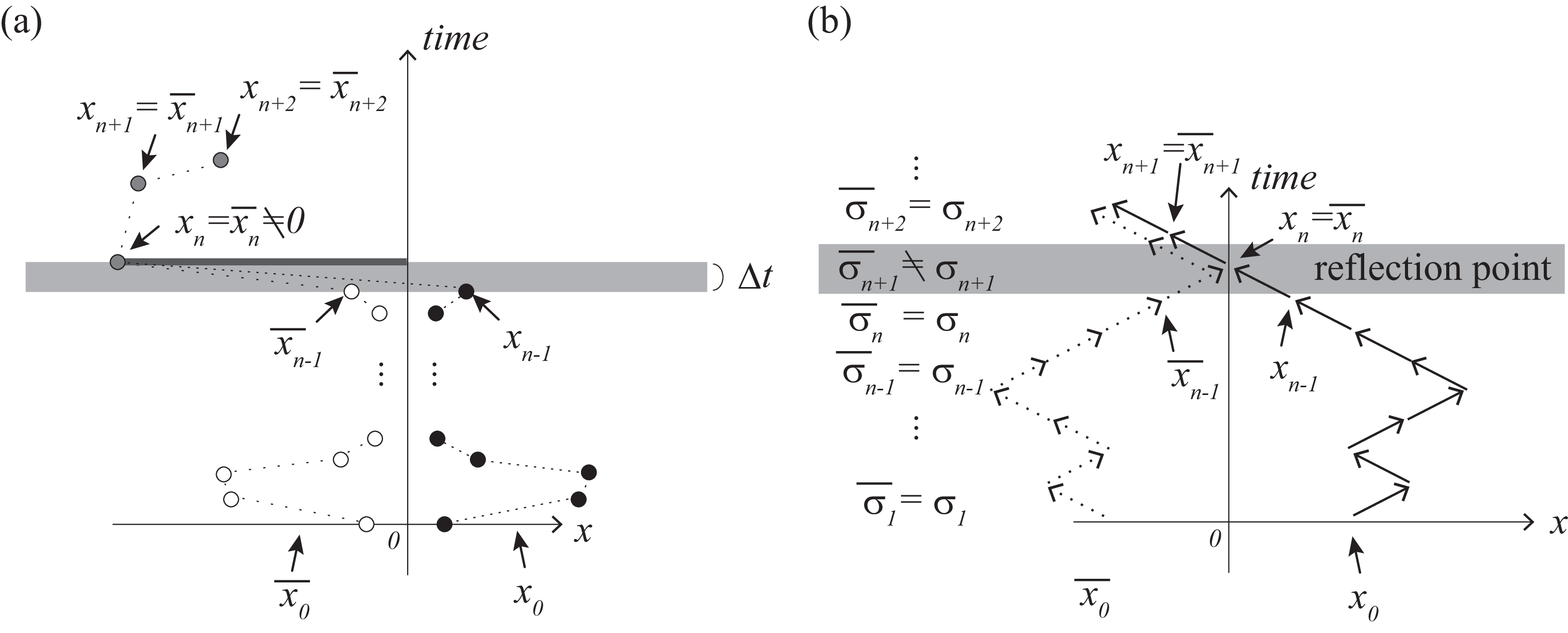}
	\caption{Schematic representations of the stochastic models to violate the MIs. (a) L\'{e}vy flight and (b) telegraph process.}
	\label{fig5}
			\vspace{0.2 cm}
\end{figure}

\subsection{Applications to other stochastic processes}

The stochastic model discussed so far can be reduced to the GLE on the ``force balance", where energetic quantities are relatively tractable.
On the other hand, there are some models for which the notion of the force balance is not so trivial.
Nonetheless, apart from energetic interpretation, we will show the consistency of the path probability ratio or the action difference as an MI violation indicator in such stochastic models.

A first example is subdiffusive continuous-time random walk (CTRW), where the MIs have been known to be available~\cite{PhysicaA_Metzler_2000}.
As a single-trajectory description, the subdiffusive CTRW can be formulated with subordination~\cite{Chaos_Grenflo_2007,EPJ_Grenflo_2011,PRE_Chechkin_2021,PCCP_Metzler_2014}.
The number of steps is counted on the subordination axis, along which the Langevin equation is given analogously.
Real elapsed time is then converted by incorporating a waiting time into each step.
Basic simple modeling decouples the waiting time with a single-step jump, which allows one to readily describe the path probability based on the Onsager-Machlup approaches.
Indeed, the probability ratio between the original and the reflected dead trajectories is specified as
\begin{eqnarray}
\frac{P_D [\{ x_{i},\Delta t_i \}_{0,n}^{N}]}{P_{\overline{D}} [\{ \overline{x}_{i},\Delta t_i \}_{0,n}^{N}]} 
&=&
\frac{\prod_{i=1}^N \psi_i(\Delta t_i) P_D [\{ x_{i} \}_{0,n}^{N}]}{\prod_{i=1}^N \psi_i(\Delta t_i) P_{\overline{D}} [\{ \overline{x}_{i} \}_{0,n}^{N}]}
=
1,~~~~
\end{eqnarray}
where $i$ denotes a step index, and $\psi_i(\Delta t_i)$ is the probability distribution of the waiting time $\Delta t_i$ at $i$-th step. 
The ratio of the path probability has no off-diagonal component in the subordination with $i$, i.e., $\Lambda[\{  \Delta x_i \}_{1,n}^N]=0$.
Thus, the MIs are found to hold.
Now, let us look back at the above applied conditions, where the Markov-process description on the subordination is crucial to enable the MIs to work.
This tells us the MIs are also utilized to investigate the first passage issues in the diffusing-diffusivity model with the stochastic diffusion constants~\cite{JPhysA_Sposini_2019}, which is formulated on the subordination~\cite{PRX_Chechkin_2017,JPhysA_Sposini_2019}.
Incidentally, the finite variances of the noises are important to identify not only the first passage, but also the first arrival at the absorbing boundary $x(\tau)=0$, as seen in a subsequent example.

In L\'{e}vy flights, steep lengths are drawn from a heavy-tailed distribution (here with infinite variance).
It has been known that MIs do not work for L\'{e}vy flights~\cite{JPhysA_Checkin_2003,AnomalousTransport_Checkin_2008} even though there is no memory effect.
Let us consider the free L\'{e}vy flight written with renewal processes, where each of the signed jump lengths $z$ is independently distributed according to the PDF $P(z)=e^{-\psi(z)}$ with symmetry $\psi(z)=\psi(-z)$.
The renewal processes might let us think that the reflection does not alter the actions, i.e., $I[\{  \Delta x_i\}_{1,n}^N]-I[\{  \Delta \overline{x}_i\}_{1,n}^N]=0$, but this is not correct. 
A simple long jump at the first passage time $\tau=n\Delta t$ may straddle the real and the image domains because of the diverging variance (see Fig.~\ref{fig5}\,(a)).
This implies we can find the well-defined first ''passage" time, but not the first ''arrival" time, where the limit $x_n \rightarrow 0$ is no longer valid even as $\Delta t\rightarrow +0$.
Hence, in general, $I[\{  \Delta x_i\}_{1,n}^N]-I[\{  \Delta \overline{x}_i\}_{1,n}^N]= \psi(x_{n}-x_{n-1})-\psi(x_{n}-\overline{x}_{n-1})\neq 0$, which indicates the MI violation.

Telegraph or Poisson-Kac processes also form one of the diverse branches in the stochastic modelings~\cite{RockyMtJMath_Kac_1974,PRE_Nizama_2024,Risks_Pogorui_2021,PRE_Sandev_2024}.
In the basic one-dimensional system, 
the left- or rightward motions are retained until the traveling directions are stochastically inverted according to Poisson processes.
Let us here see the path probability ratio in the unbiased Telegraph process with constant speed.
The construction of the reflected path (see Eq.~(\ref{x_dead}) and Fig.~\ref{fig1}) requires either one of the dead or the reflected walkers to turn around at the absorbing boundary to create a pair of walkers. 
This makes the path probability ratio take nonunity as
\begin{eqnarray}
\frac{P_D [\{ x_{i} \}_{0,n}^{N}]}{P_{\overline{D}} [\{ \overline{x}_{i} \}_{0,n}^{N}]} 
&=&
\frac{\prod_{i=1}^N (\lambda \Delta t)^{\sigma_i}(1-\lambda \Delta t)^{1-\sigma_i} }{\prod_{i=1}^N (\lambda \Delta t)^{\overline{\sigma}_i}(1-\lambda \Delta t)^{1-\overline{\sigma}_i} }
\nonumber \\
&=&
\frac{ (\lambda \Delta t)^{\sigma_{n+1}}(1-\lambda \Delta t)^{1-\sigma_{n+1}} }{ (\lambda \Delta t)^{1-\sigma_{n+1}}(1-\lambda \Delta t)^{\sigma_{n+1}} } \neq 1,
\end{eqnarray}
where $\sigma_i=1$ is drawn to turn around with the probability $\lambda \Delta t$ at $i$-th step, otherwise $\sigma_i=0$.
In parallel, the counterpart in the reflected path is also denoted by $\overline{\sigma}_i$.
While $\sigma_i=\overline{\sigma}_i$ for $i \neq n+1$, turning around at $n$ implies $\sigma_{n+1}+\overline{\sigma}_{n+1}=1$ (e.g., see Fig.~\ref{fig5}\,(b) without a turn ($\sigma_{n+1}=0$) and with a turn ($\overline{\sigma}_{n+1}=1$) at the reflection point).
Thus, the probability at $n+1$ is not cancelled out between the numerator and the denominator in the probability ratio, indicating the MI violation.

\subsection{The family of Gallavotti-Cohen symmetry}

Recent theoretical works~\cite{JPhysA_Sarmiento_2025,PRL_Gingrich_2017} have pointed out an interesting connection between the fluctuation theorem and certain symmetry properties of the first passage time.
In these works, the fluctuation theorem is represented in the generating function form known as the Gallavotti-Cohen symmetry.
In our case, taking the ensemble average after multiplying $e^{2 \Lambda [\{  \Delta x_i\}_{1,n}^N]} P_D [\{ x_{i} \}_{0,n}^{N}] =P_{\overline{D}} [\{ \overline{x}_{i} \}_{0,n}^N]$ (Eq.~(\ref{path_ratio})) with $\delta (x-x_N)e^{2(\eta-1) \Lambda[\{ \Delta x_{i} \}_{1,n}^{N}]}$ and using $\Lambda [\{  \Delta \overline{x}_i\}_{1,n}^N]=-\Lambda [\{  \Delta x_i\}_{1,n}^N]$ along essentially the same procedures in ref.~\cite{JCP_Gaspard_2004}
, we encounter the family of the Gallavotti-Cohen symmetry:
\begin{eqnarray}
\frac{ P_D(x,N|x_0)}{ P_{{\overline{D}}}(x,N|\overline{x}_0)} 
=
\frac{\left< e^{2(1-\eta) \Lambda} \right>_{\overline{D},x}}{\left< e^{2\eta \Lambda} \right>_{D,x}},
\label{vi_IM_eta}
\end{eqnarray}
where $\eta$ is a parameter.
Equation~(\ref{vi_IM_eta}) corresponds to a generating-function form of Eq.~(\ref{vi_IM}).

\subsection{Towards nonGaussian formulated with mutual information}

Lastly, we discuss the path correlations $\Lambda$ quantified by mutual information, while $\Lambda$ has been seen to be associated with the heat in the stochastic processes compatible with the GLE.
Recall that the heat notion was introduced by virtue of the force balance in the main text, but the force balance is not always obvious.
Nonetheless, the path correlations in the MIs are quantified with
\begin{eqnarray}
2 \Lambda [\{ x_{i} \}_{0,n}^{N}]
=
i_{\overline D} [\{ {\overline x}_{i} \}_{0,n}^{N}]
-
i_D[\{ x_{i} \}_{0,n}^{N}],
\label{Lambda_MutualInfo}
\end{eqnarray}
where
\begin{eqnarray}
i_D[\{ x_{i} \}_{0,n}^{N}]\coloneqq \ln{\left( \frac{P_D[\{ x_{i} \}_{0,n}^{N}]}{ P_D[\{ x_{i} \}_0^{n}]P_D[\{ x_{i} \}_{n}^{N}]} \right)}
\label{iD_def}
\end{eqnarray}
 is mutual information between pre- and post-first passage paths in a dead path.
In the same vein, $i_{\overline D} [\{ {\overline x}_{i} \}_{0,n}^{N}]\coloneqq \ln{\left( P_{\overline D}[\{ \overline{x}_{i} \}_{0,n}^{N}]/P_{\overline D}[\{ \overline{x}_{i} \}_0^{n}]P_{\overline D}[\{ \overline{x}_{i} \}_{n}^{N}] \right)}$ is defined for the reflected path.

The quantification by Eq.~(\ref{Lambda_MutualInfo}) is not dependent on the detailed structures of the stochastic processes, i.e., the mathematical formulation with Eq.~(\ref{Lambda_MutualInfo}) can be employed not only in Gaussian, but also in nonGaussian processes.
This gives a potential direction in the systems formulated by the probabilistic descriptions, but the thermodynamic or physical perspectives are left for future issues.

\section{Concluding Remarks}
\label{Conclusion_sec}

In summary, we have shown that the MIs fail in non-Markovian processes due to the lack of reflection symmetry. 
The symmetry is broken by the memory effect, and the degree of asymmetry manifests in the history-dependent conditional probability.
For the stochastic processes reducible to the GLE with the FDR, the deviation from the MI result can be quantified by the heat, whose statistics mirror characteristic features of the process.

This article has not discussed entropy production to complete the stochastic thermodynamics.
Formulating the entropy production by embedding the heat $2\Lambda$ through, e.g., Eqs.~(\ref{IFT_Lambda}),\,(\ref{Lambda_Clausius}), should be a necessary step to elucidate memory effects in non-Markovian processes under absorbing boundaries.

It may be interesting to investigate the problem using, e.g., the Rouse polymer model~\cite{DoiEdwards}, where the interaction between the system (in the present context, a tagged monomer in the polymer) and the environment (the rest of the polymer) can be explicitly analyzed~\cite{Panja_2010,PRE_Sakaue_2013,Maes_2013,PRE_Saito_Sakaue_2015,Vandebroek_2017,PRE_Saito_2022}.  
We expect that elucidating the thermodynamic aspect of such a system may provide valuable insights into the first passage problem in non-Markovian processes.

\backmatter

\if0
\bmhead{Supplementary information}

If your article has accompanying supplementary file/s please state so here. 

Authors reporting data from electrophoretic gels and blots should supply the full unprocessed scans for key as part of their Supplementary information. This may be requested by the editorial team/s if it is missing.

Please refer to Journal-level guidance for any specific requirements.
\fi

\bmhead{Acknowledgements}

The authors thank Prof. K. Seki at National Institute of Advanced Industrial Science and Technology 
for fruitful discussions.

\section*{Statements and Declarations}

\bmhead{Conflict of interest/Competing interests}

The authors have no relevant financial or non-financial interests to disclose.

\bmhead{Funding}

This work was supported by JSPS KAKENHI (Grant No.JP23H00369, JP24K00602 and JP26K07037).

\bmhead{Data availability}

All the numerically generated data are shown in the manuscript.

\if0
\section*{Declarations}

Some journals require declarations to be submitted in a standardised format. Please check the Instructions for Authors of the journal to which you are submitting to see if you need to complete this section. If yes, your manuscript must contain the following sections under the heading `Declarations':

\begin{itemize}
\item Funding
\item Conflict of interest/Competing interests (check journal-specific guidelines for which heading to use)
\item Ethics approval and consent to participate
\item Consent for publication
\item Data availability 
\item Materials availability
\item Code availability 
\item Author contribution
\end{itemize}

\noindent
If any of the sections are not relevant to your manuscript, please include the heading and write `Not applicable' for that section. 

%%===================================================%%
%% For presentation purpose, we have included        %%
%% \bigskip command. Please ignore this.             %%
%%===================================================%%
\bigskip
\begin{flushleft}%
Editorial Policies for:

\bigskip\noindent
Springer journals and proceedings: \url{https://www.springer.com/gp/editorial-policies}

\bigskip\noindent
Nature Portfolio journals: \url{https://www.nature.com/nature-research/editorial-policies}

\bigskip\noindent
\textit{Scientific Reports}: \url{https://www.nature.com/srep/journal-policies/editorial-policies}

\bigskip\noindent
BMC journals: \url{https://www.biomedcentral.com/getpublished/editorial-policies}
\end{flushleft}
\fi

%% \begin{appendices}
%% \end{appendices}
\section*{Appendix}

\renewcommand{\thesection}{\Alph{section}}
\setcounter{section}{0}

This appendix is organized as follows.  
Section~\ref{sec:0} provides the continuous-form analyses with the GLE.
In Sec.~\ref{sec:1}, we present a derivation of the fluctuation theorem for $\Lambda$ (Eq.~\eqref{DF_2}) and supplement the relevant analyses.
Section~\ref{sec:2} discusses an equality for the survival probability, which is obtained by integrating Eq.~\eqref{DF_2} over $\Lambda$ and $x\, ( \ge 0)$.
Section~\ref{sec:3} describes a summary of the numerical simulation method and some supplemental numerical results related to the discussion in the main text.
In Sec.~\ref{Class_FT}, we supplement the classifications of entropic functionals associated with the fluctuation theorem.

\section{Continuous forms with GLE}
\label{sec:0}

Some readers might be used to continuous representations in the path probability with the Langevin equation.
The following continuous formalism is compatible with the discrete one in the main text.
We here consider a one-dimensional overdamped Langevin equation corresponding to the discrete Eq.~(\ref{EM_vel}):
\begin{eqnarray}
\frac{dx_t}{dt}
=\xi_t,
\label{LangevinEq_continuum}
\end{eqnarray}
where $\xi_t$ is Gaussian noises with zero mean $\left< \xi_t \right>=0$ and the covariance $\left< \xi_t\xi_{t'} \right>=C_{tt'}$.
A temporal sequence of the noises $\{ \xi_{s} \}_0^t$, whose probability is dictated by
\begin{eqnarray}
P_0^{(\xi)}[\{ \xi_s \}_0^t]
\sim
\exp{\left( -\frac{1}{2}\int_0^t ds\int_0^tds'\,C_{ss'}^{-1} \xi_s \xi_{s'} \right)},
\label{Pro_noise_Lang}
\end{eqnarray}
where $C_{ss'}^{-1}$ is the Green function satisfying $\int_0^tds'\,C_{ss'}^{-1}C_{s's''}=\delta(s-s'')$ and symmetric with $C_{ss'}^{-1}=C_{s's}^{-1}$.

Equation~(\ref{Pro_noise_Lang}) determines the path trajectory $\{ x_s \}_0^t$ and its probability given an initial position $x_0$.
%% To see the connection with the heat, a functional of the velocity $\dot{x}_t\coloneqq dx_t/dt$ is suitable, and 
The Onsager-Machlup approaches with Eq.~(\ref{LangevinEq_continuum}) transform Eq.~(\ref{Pro_noise_Lang}) into
\begin{eqnarray}
P_0[\{ x_s \}_0^t]
\sim
\exp{\left( -\frac{1}{2}\int_0^t ds\int_0^tds'\,C_{ss'}^{-1}\dot{x}_s\dot{x}_{s'} \right)},
\label{P0_OM}
\end{eqnarray}
where $\dot{x}_t\coloneqq dx_t/dt$.
Note that the Jacobian ${\rm det}[\delta \xi_{s}/\delta \dot{x}_{s'}]=const.$ drops for notational conveniences.
%%  because it turns out to be irrelevant to observe a ratio of the path probabilities.
In addition, $P_D[\{ x_s \}_0^t]$, $P_{\overline{D}}[\{ x_s \}_0^t]$, and $P_S[\{ x_s \}_0^t]$ are extracted by inserting, e.g., dead-path filter $\left( 1-\prod_{s=0}^t \Theta^+(x_s)  \right)$ into the right hand side of Eq.~(\ref{P0_OM}) as around Eq.~(\ref{P_D_path}), but we are restricted into the dead path, which allows us not to explicitly express the filters.

Let the first passage time be $\tau$, where a reflected path is drawn along $\overline{x}_s=-x_s$ for $0 \leq s \leq \tau$ and $\overline{x}_s=x_s$ for $\tau \leq s \leq t$ (see Eq.~(\ref{x_dead}) and Fig.~\ref{fig1}).
A ratio of $P_D[\{ x_s \}_{0,\tau}^t] \sim \exp{\left( -\frac{1}{2}\int_0^t ds\int_0^tds'\,C_{ss'}^{-1}\dot{x}_s\dot{x}_{s'} \right)}$ with $P_{\overline{D}}[\{ \overline{x}_s \}_{0,\tau}^t] \sim \exp{\left( -\frac{1}{2}\int_0^t ds\int_0^tds'\,C_{ss'}^{-1}\dot{\overline{x}}_s\dot{\overline{x}}_{s'} \right)}$ quantifies the violation component of the MIs:
\begin{eqnarray}
\frac{
P_D[\{ x_s \}_{0,\tau}^t]
}{
P_{\overline{D}}[\{ \overline{x}_s \}_{0,\tau}^t]
}
=
\exp{[
-2\Lambda[\{ x_s \}_{0,\tau}^t]
]}.
\end{eqnarray}
Then, $\Lambda[\{ x_s \}_{0,\tau}^t]=\int_\tau^t ds\int_0^\tau ds'\,C_{ss'}^{-1}\dot{x}_s\dot{x}_{s'}$ is found to be identical to Eq.~(\ref{Lambda_def_Lang}) with $\Gamma_{ss'} \coloneqq C_{ss'}^{-1}$.

To see the connection of $\Lambda[\{ x_s \}_{0,\tau}^t]$ to the heat, we first discover the equation of motion conditional on $\{ x_s\}_0^\tau$.
Completing the square of the argument of the dead path probability with variables $\{ x_s\}_\tau^t$  at a fixed $\{ x_s\}_0^\tau$ rewrites it to
\begin{eqnarray}
&&
P_D[\{ x_s \}_{0,\tau}^t]
\nonumber \\
&\sim&
\exp{}
\Biggl[ -\frac{1}{2}\int_0^t ds\int_0^tds'\, \Gamma_{ss'}  \dot{x}_s \dot{x}_{s'} \Biggr]
\nonumber \\
&=&
\exp{}
\Biggl[ -\frac{1}{2}\int_0^\tau ds\int_0^\tau ds'\, 
\nonumber \\
&& \times
\left(\Gamma_{ss'}  - \int_\tau^t du\int_\tau^t du'\, \Gamma_{su} (\Gamma^{\rm II})_{uu'}^{-1} \Gamma_{s'u'} \right)
\dot{x}_s \dot{x}_{s'} \Biggr]
\nonumber \\
&&
\times \exp{}
\Biggl[ -\frac{1}{2}\int_\tau^t ds\int_\tau^tds'\, \Gamma_{ss'}
\left( \dot{x}_s -\int_\tau^t du\, (\Gamma^{\rm II})^{-1}_{su}f_u  \right)
\nonumber \\
&& \times
\left( \dot{x}_{s'} -\int_\tau^t du'\, (\Gamma^{\rm II})^{-1}_{s'u'}f_{u'}  \right)
\Biggr],
\label{PD_completeSq}
\end{eqnarray}
where the ``memory force" is $f_s=-\int_0^\tau dw\,\Gamma_{sw}\dot{x}_w$, and $(\Gamma^{\rm II})^{-1}_{su}$ is defined such that  $\int_\tau^t du\, (\Gamma^{\rm II})^{-1}_{su}\Gamma_{us'}=\delta(s-s')$.
A notational caveat is that while in the discrete description of the main text, we have introduced $m\times m$ submatrix $\Gamma_{i-n,j-n}^{\mathrm{II}}=\Gamma_{i,j}$ for $i,\,j \in [n+1,N]$, here, we define the kernel $\Gamma^{\rm II}_{us} \coloneqq \Gamma_{us}$ for $u,s\in [\tau,t]$ (see the top of Eqs.~(\ref{blcok_original_matrix})).
In Eq.~(\ref{PD_completeSq}), while the first exponential factor in the last equation is kept unchanged at a fixed first passage path $\{ \dot{x}_s \}_0^\tau$, the last exponential provides the functional conditional on $\{ \dot{x}_s \}_0^\tau$ through the memory force.
The argument in the last exponential leads us to introduce a sequence of Gaussian noises $\{ \xi_s^{\mathrm{II}} \}_\tau^t$ such that
\begin{eqnarray}
\dot{x}_s -\int_\tau^t du\, (\Gamma^{\rm II})^{-1}_{su}f_u = \xi_s^{\mathrm{II}},
\label{EffectiveEOM_vel}
\end{eqnarray}
which simply rewrites Eq.~(\ref{PD_completeSq}) as $P_D[\{ x_s \}_{0,\tau}^t] \sim P_D^{(\xi^{\mathrm{II}})}[\{ \xi_s^{\mathrm{II}} \}_{0,\tau}^t] \sim \exp{} [ -(1/2)\int_\tau^t ds\int_\tau^tds'\, \Gamma_{ss'} \xi_s^{\mathrm{II}}\xi_{s'}^{\mathrm{II}} ]$ for a fixed $\{ \dot{x}_s \}_0^\tau$.
Furthermore, operating $\int_\tau^t ds\, \Gamma_{ws} (\cdot) $ on both the sides of Eq.~(\ref{EffectiveEOM_vel}), we have
\begin{eqnarray}
\int_\tau^t ds\, \Gamma_{ws} \dot{x}_s = f_w +\zeta_w,
\end{eqnarray}
where $\zeta_w \coloneqq \int_\tau^t ds\, \Gamma_{ws} \xi_s^{\mathrm{II}}$.
By putting $w=t$, the force balance implies the GLE under the memory force:
\begin{eqnarray}
\int_\tau^t ds\, \Gamma_{ts} \dot{x}_s = f_t +\zeta_t.
\label{ForceBalance_GLE}
\end{eqnarray}
Recall the heat defined with the Langevin equation by Sekimoto~\cite{Sekimoto_book}.
Analogously, the heat for the GLE is defined as
\begin{eqnarray}
Q
=
\int_\tau^t
d'Q
&=&
\int_\tau^t ds\,
\left( 
-\int_\tau^s du\, \Gamma_{su} \dot{x}_u +\zeta_s
\right)
\frac{dx_s}{ds}.
\label{GLE_heat_def}
\end{eqnarray}
Note that the Markovian limit with $\Gamma_{ss'}=2\gamma \delta (s-s')$ turns Eq.~(\ref{GLE_heat_def}) into $Q=\int_\tau^t ds\,\left( -\gamma \dot{x}_s +\zeta_s\right) \circ \frac{dx_s}{ds}$ with $\circ$ denoting Stratonovich multiplication, which is ensured by starting with interpreting the multiplication in Eq.~(\ref{Pro_noise_Lang}) as $\circ$.
Although the Stratonovich multiplication is generally critical in the heat definition, it does not matter in the present MI systems since the derivation through Eqs.~(\ref{Pro_noise_Lang})-(\ref{GLE_heat_def}) do not have the multiplicative term to distinguish the Stratonovich from the Ito multiplication.

A final step is to apply the force balance (Eq.~(\ref{ForceBalance_GLE})) to the heat definition (Eq.~(\ref{GLE_heat_def})):
\begin{eqnarray}
Q
&=&
-\int_\tau^t ds\,
f_s \frac{dx_s}{ds}
\nonumber \\
&=&
\int_\tau^t ds\,
\int_0^\tau ds'\,\Gamma_{ss'}\dot{x}_{s'}
 \dot{x}_s
=\Lambda [\{ x_s \}_{0,\tau}^t].
\end{eqnarray}
We eventually find the heat amounts to $\Lambda [\{ x_s \}_{0,\tau}^t]$.

\begin{figure}[t]
	\centering
	\includegraphics[width=1.0\textwidth]{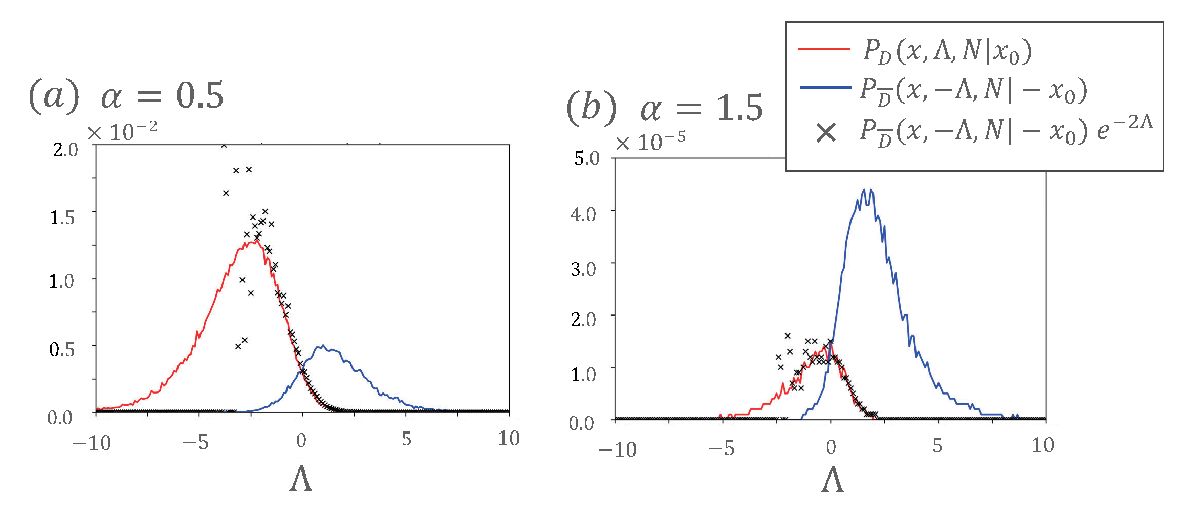}
	\caption{Numerical plots of the key factors appearing in Eq.~(\ref{DF_2}) for (a) subdiffusive fBm ($\alpha=0.5$, $x=5$, $x_0=\sqrt{300^{\alpha}}$, $N=1\times 10^3$) and (b) superdiffusive fBm ($\alpha=1.5$, $x=100$, $x_0=\sqrt{300^{\alpha}}$, $N=1\times 10^3$). Red or blue solid line represents $P_D(x,\Lambda,N|x_0)$ or, $P_{\overline{D}}(x,-\Lambda,N|\overline{x}_0)$, respectively. Black cross symbols indicate $P_{\overline{D}}(x,-\Lambda,N|\overline{x}_0)e^{-2\Lambda}$. 
	}
	\label{figS1}
			\vspace{0.2 cm}
\end{figure}

\section{Fluctuation theorem for $\Lambda$}
\label{sec:1}

This section presents derivations around Eqs.~(\ref{ft_Lambda})-(\ref{vi_IM}) together with the definitions of the conditional averages such as $\langle (\cdot) \rangle_{D}$ and $\langle (\cdot) \rangle_{D,x}$.

The probability $P_D(x,\Lambda, N|x_0) $ is obtained by extracting the paths with $\Lambda[\{ \Delta x_i\}_{1,n}^N]=\Lambda$ and $x_N=x$ through an insertion of Dirac's delta into the path integral of Eq.~(\ref{P}):
\begin{eqnarray}
P_D(x,\Lambda, N|x_0) 
&=& 
\sum_{n=1}^N \prod_{i=1}^n \int_0^{\infty} dx_i \prod_{i'=n+1}^N \int_{-\infty}^{\infty} dx_{i'}
\nonumber \\
&& \times
  \  P_0[\{ x_{i} \}_{0,n}^{N}]  \delta (x -x_N) 
\nonumber \\
&& \times
\delta (\Lambda -\Lambda[\{ \Delta x_i\}_{1,n}^N] ),
\label{DF_2_cal}
\end{eqnarray}
where the integral domains are divided into $\int_0^\infty$ and $\int_{-\infty}^{\infty}$ by the first passage time $n$.
Technically, an introduction of $\Theta^+(x_i)$ (see the definition after Eq.~(\ref{P_D_path})) gives a different form without sorting by cases:
\begin{eqnarray}
P_D(x,\Lambda, N|x_0)
&=& 
\prod_{i=1}^N \int_{-\infty}^{\infty} dx_i
\nonumber \\
&&
\times \left(1-\prod_{i'=1}^N \Theta^+(x_{i'})\right)
P_0[\{ x_{i} \}_{0,n}^{N}] 
\nonumber \\
&& \times \delta (x -x_N) \delta (\Lambda -\Lambda[\{ \Delta x_i\}_{1,n}^N] )
\nonumber \\
&=& 
\prod_{i=1}^N \int_{-\infty}^{\infty} dx_i
  \  P_D[\{ x_{i} \}_{0,n}^{N}] 
\nonumber \\
&& \times \delta (x -x_N) \delta (\Lambda -\Lambda[\{ \Delta x_i\}_{1,n}^N] ).
\end{eqnarray}
By using Eq.~(\ref{path_ratio}) and performing the variable transformation: $\{ \Delta x_i\}_{1,n}^N] \rightarrow \{ \Delta \overline{x}_i\}_{1,n}^N$, with the parity $\Lambda[\{ \Delta \overline{x}_i\}_{1,n}^N]=-\Lambda[\{ \Delta x_i\}_{1,n}^N]$, we arrive at Eq.~(\ref{DF_2}) as
\begin{eqnarray}
P_D(x,\Lambda, N|x_0) 
&=& 
\prod_{i=1}^N \int_{-\infty}^{\infty} d\overline{x}_i
  \  P_{\overline{D}}[\{ \overline{x}_{i} \}_{0,n}^{N}] e^{2 \Lambda [\{  \Delta \overline{x}_i\}_{1,n}^N]}
\nonumber \\
&& \times \delta (x -\overline{x}_N) \delta (\Lambda +\Lambda[\{ \Delta \overline{x}_i\}_{1,n}^N] ).
\nonumber \\
&=& 
P_{\overline{D}}(x,-\Lambda, N|x_0) e^{-2 \Lambda}.
\end{eqnarray}
Note that the probability conditional on $(x,\Lambda)$ is defined as
\begin{eqnarray}
P_Z(x,\Lambda,N|z_0)
&\coloneqq& \prod_{i=1}^N \int_{-\infty}^{\infty}dz_i\, \delta (x -x_N)
\nonumber \\
&& \times \delta (\Lambda -\Lambda[\{ \Delta z_i \}_{1,n}^N ])P_Z[\{ z_i \}_{0,n}^N],
\end{eqnarray}
where $Z=D, \overline{D}$ inputs $z_i=x_i, \overline{x}_i$, respectively.

In Fig.~\ref{figS1}, we show a numerical demonstration of the above relation.
 The collapse of $P_{\overline{D}}(x,-\Lambda,N|\overline{x}_0)e^{-2\Lambda}$ onto $P_D(x,\Lambda,N|x_0)$ (except for the skirt due to deteriorating sampling efficiency) supports Eq.~(\ref{DF_2}).

In the same vein, we find Eq.~(\ref{ft_Lambda}) by taking $\delta (x-x_N)$ off in Eq.~(\ref{DF_2_cal}).

Explicitly, the ensemble average over the dead or the reflected path is defined as
\begin{eqnarray}
\langle (\cdot) \rangle_{Z} \coloneqq P_Z(N|z_0)^{-1} \prod_{i=1}^N \int_{-\infty}^{+\infty} dz_i\, (\cdot) P_Z[\{ z_{i} \}_0^{N}].
\end{eqnarray}
Bear in mind that $\langle (\cdot) \rangle_{Z}$ is normalized by the probability $P_Z(N|z_0)$ at the terminal time $N$.
This definition provides the same form of the conventional integral fluctuation theorem:
\begin{eqnarray}
\left< e^{2\Lambda} \right>_D=1.
\end{eqnarray}
Similarly, the average conditional on $x$ at $N$ is defined as
\begin{eqnarray}
\langle (\cdot) \rangle_{Z,x}\coloneqq P_Z(x,N|z_0)^{-1} \int_{-\infty}^{\infty}d\Lambda\,(\cdot)P_Z(x,\Lambda,N|z_0)
\label{def_aveZx}
\end{eqnarray}
with
$
P_Z(x,N|z_0)\coloneqq \int_{-\infty}^{\infty}d\Lambda\,P_Z(x,\Lambda, N|z_0).
\label{def_PZx}
$
Using Eqs.~(\ref{def_aveZx}), the analogous calculations lead us to Eq.~(\ref{vi_IM}).

\vspace{1cm}
\section{Survival probability}
\label{sec:2}

One can also obtain the following expression for the survival probability $P_S(t,x_0) \coloneqq \int_0 ^{\infty} dx \ P_S(x,t|x_0)$ by integration over the physical domain $x \ge 0$;
\begin{eqnarray}
P_S(t,x_0) = \frac{ 1 + \mathrm{erf}( x_0/\sigma_t )  }{2}-   \frac{ 1 - \mathrm{erf}( x_0/\sigma_t )  }{2  \langle \exp{(2 \Lambda)}\rangle_{D, +} },
\label{S_t}
\end{eqnarray}
where $\sigma_t \coloneqq  [\int_{-\infty}^{\infty} dx \ (x-x_0)^2 P_0(x,t|x_0)]^{1/2}$ is the root mean-square displacement of the unrestricted walkers and the average of the dead paths conditional on the positive domain is defined as $\langle  (\cdot) \rangle_{D, +} \coloneqq \int_{-\infty}^\infty d\Lambda\,(\cdot) \int_0^{\infty} dx \  P_D(x,\Lambda,t|x_0)/[ \int_0 ^{\infty} dx \  P_D(x,t|x_0)]$.

Before deriving Eq.~\eqref{S_t}, we define the following quantities through the integral of the probability distribution of the dead walkers $P_D(x,t|x_0)$;
\begin{eqnarray}
H_D^{+}(t,x_0) \coloneqq \int_0^{\infty}dx\,P_D(x,t|x_0), \nonumber \\
H_D^{-}(t,x_0) \coloneqq \int_{-\infty}^{0}dx\,P_D(x,t|x_0), 
\end{eqnarray}
where the superscript $+$ or $-$ indicates the integral domain. Similarly, we define $H_{\overline{D}}^{+}(t,\overline{x}_0)$, $H_{\overline{D}}^{-}(t,\overline{x}_0)$ from $P_{\overline{D}}(x,t|\overline{x}_0)$ and  $H_{0}^{+}(t,x_0)$, $H_{0}^{-}(t,x_0)$ from $P_0(x,t|x_0)$.
One can verify the following relations:
\begin{eqnarray}
H_{\overline{D}}^{+}(t,\overline{x}_0) = H_{0}^{+}(t,-x_0) = H_{0}^{-}(t,x_0),
\label{H-relation}
\end{eqnarray}
where the first equality follows from Eq.~\eqref{P_barD_P_0-} and the second equality is readily obtained by the change of the integration variable $x \rightarrow -x$. 

To derive Eq.~\eqref{S_t}, we start from the relation $ P_D(x,N|x_0) =  P_{\overline{D}}(x,N|\overline{x}_0) \langle  \exp{(2 \Lambda)} \rangle_{\overline{D},x}$ or $P_D(x,N|x_0) \langle  \exp{(2 \Lambda)} \rangle_{D,x} =  P_{\overline{D}}(x,N|\overline{x}_0)$ from Eq.~\eqref{vi_IM}. Integrating both sides of the equation over the physical domain $x \ge 0$, we find
\begin{eqnarray}
\frac{H_D^{+}(N,x_0)}{H_{\overline{D}}^{+}(N,x_0)} = \langle \exp{(2 \Lambda)}\rangle_{\overline{D}, +} =  \langle \exp{(2 \Lambda)}\rangle_{D, +}^{-1},
\label{H_ratio}
\end{eqnarray}
where the average of the reflected paths conditional on the positive domain is defined likewise:
$\langle  (\cdot) \rangle_{\overline{D}, +} \coloneqq \int_{-\infty}^\infty d\Lambda\,(\cdot) \int_0^{\infty} dx \  P_{\overline{D}}(x,\Lambda,N|\overline{x}_0)/[ \int_0 ^{\infty} dx \  P_{\overline{D}}(x,N|\overline{x}_0)]$.
Using Eqs.~\eqref{H-relation} and~\eqref{H_ratio} and the definition of the survival probability $P_S(t,x_0)  \coloneqq \int_0 ^{\infty} dx \ P(x,t|x_0) =  H_{0}^{+}(t,x_0) - H_D^{+}(t,x_0)$, we find
\begin{eqnarray}
P_S(t,x_0) &=& H_0^{+}(t,x_0) -  \langle \exp{(2 \Lambda)}\rangle_{\overline{D}, +} H_0^{-}(t,x_0) \nonumber \\
&=& H_0^{+}(t,x_0) - H_0^{-}(t,x_0) / \langle \exp{(2 \Lambda)}\rangle_{D, +}.
\label{S_H}
\end{eqnarray}
For Gaussian processes, the distribution of unrestricted walkers starting from $x=x_0$ is
\begin{eqnarray}
P_0(x,t|x_0) = \frac{1}{\sqrt{2 \pi \sigma_t^2}} \exp{\left( - \frac{(x-x_0)^2}{2 \sigma_t^2} \right)}
\end{eqnarray}
and its integral over the physical (positive) or negative domain is represented with the error function $ \mathrm{erf}(x) \coloneqq (2/\sqrt{\pi}) \int_0^x e^{-y^2} dy$;
\begin{eqnarray}
H_0^{\pm}(t,x_0) = \frac{1}{2} \left( 1 \pm \mathrm{erf} \left( \frac{x_0}{\sigma_t}\right)\right) 
%H_0^{-}(t,x_0) = \frac{1}{2} \left( 1 - \mathrm{erf} \left( \frac{x_0}{\sigma_t}\right)\right)
\label{H_0}
\end{eqnarray}
with the double sign in the same order.
Combining Eqs.~\eqref{S_H} and~\eqref{H_0}, we obtain Eq.~\eqref{S_t}.

\begin{figure}[t]
	\centering
	\includegraphics[width=1.0\textwidth]{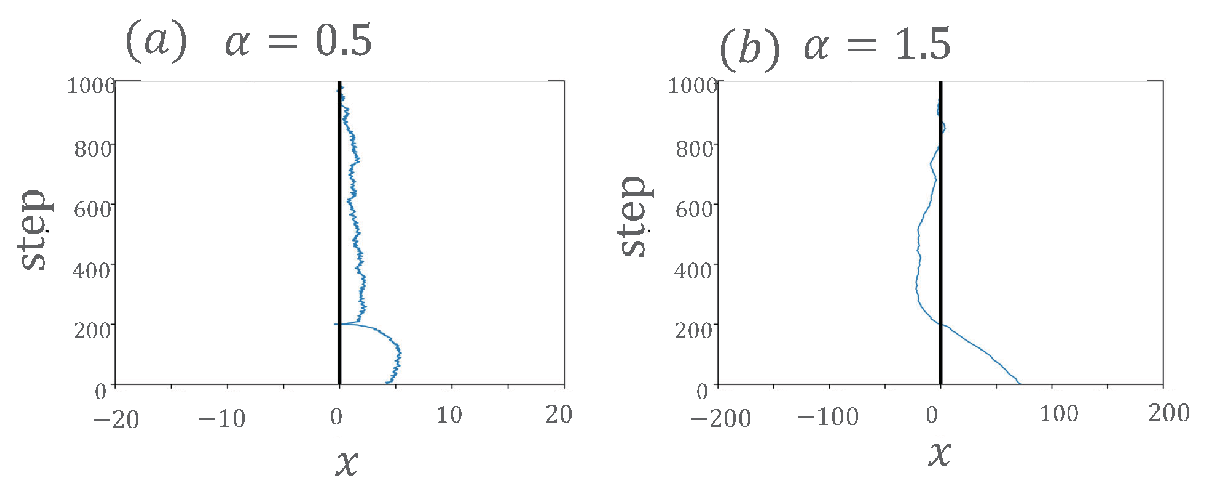}
	\caption{Sample averages of conditional paths with $x^*=0$ for (a) subdiffusive fBm ($\alpha = 0.5$) with $x_0=\sqrt{300^{\alpha}}$, $n=200$, $N=1\times 10^3$ and (b) superdiffusive fBm ($\alpha=1.5$) with $x_0=\sqrt{300^{\alpha}}$, $n=200$, $N=1\times 10^3$. The averaging is done with 71 sample paths for (a) and 30 sample paths for (b). 
	}
	\label{figS2}
			\vspace{0.2 cm}
\end{figure}

\begin{figure}[t]
	\centering
	\includegraphics[width=1.0\textwidth]{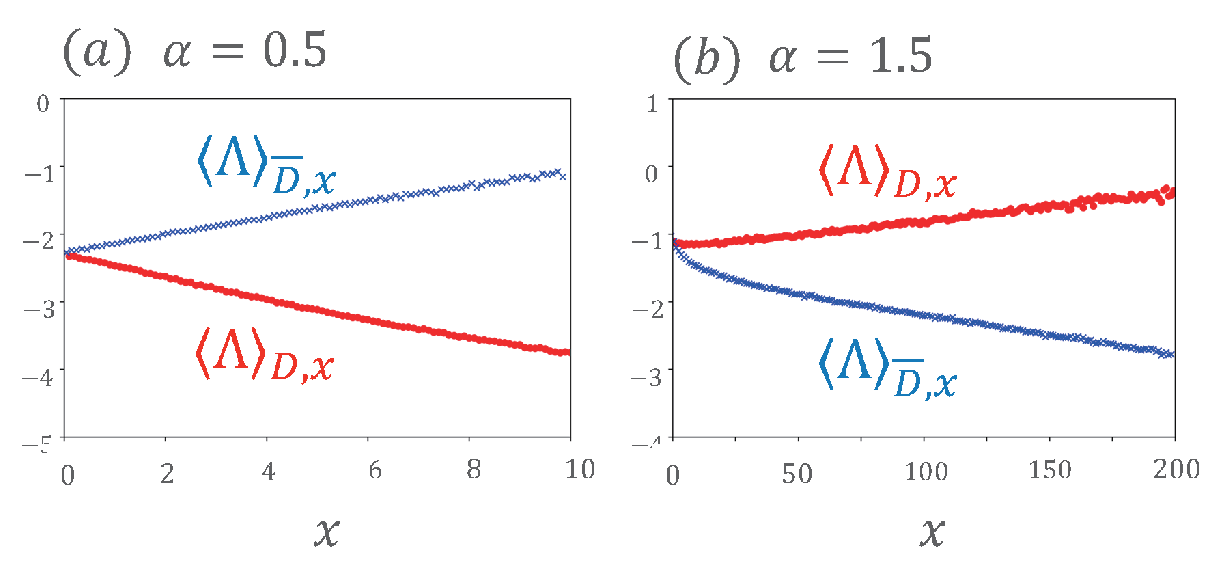}
	\caption{Average heat flow in the post-first-passage process. (a) fBm with $\alpha=0.5$ and (b) fBm with $\alpha=1.5$. The starting position is $x_0=300^{\alpha/2}$.
	Average absorbed heat $  \langle \Lambda \rangle_{D,x}$ along the dead paths (red)  and $ \langle \Lambda \rangle_{{\overline D},x}$ along the reflected paths (blue) are shown as functions of the endpoint $x\, (\ge 0)$ after $N=10^3$ steps. Note that the distribution of these heats is shown in Fig.~\ref{figS1}.} 
	\label{figS3}
			\vspace{0.2 cm}
\end{figure}

\section{Numerical simulations}
\label{sec:3}

This section notes the numerical scheme throughout the article and supplemental data.
The one-dimensional motion set off from $x_0$ is updated by iterating
\begin{eqnarray}
\frac{x_{i}-x_{i-1}}{\Delta t}=\xi_i,
\label{Numerical_EOM}
\end{eqnarray}
where $i$ denotes the number of the steps in the recursive loop with $\Delta t$ being a step size, and $\xi_i$ denotes the noise.
The limit $\Delta t \rightarrow +0$ reduces eq.~(\ref{Numerical_EOM}) into the equation of motion $dx_t/dt=\xi_t$ at a time $t\,(=i\Delta t)$ in the continuous picture. Unless otherwise stated, the numerical simulations consider dimensionless units and set $\Delta t=1$.
Incidentally, $\Delta t=1$ also reduces Eq.~(\ref{Numerical_EOM}) to the discrete form of Eq.~(\ref{EM_vel}) in the main text.

The statistics of the noises $\{ \xi_i\}_1^N$ dictate the stochastic processes.
Colored noises of the nonMarkovian fBms are generated according to the Hosking method~\cite{Hosking,Dieker_2004}.
For Markovian walkers ($\alpha=1$),  $\{ \xi_i\}_1^N$ is simply taken from independently and identically distributed Gaussian noise.

\subsection{Conditional paths}

From an ensemble of paths starting from $x=x_0$, we extract paths, which arrive at $x=0$ for the first time at $i=n$, and visit $x=x^*$ at $i=N$.
Here we call such paths {\it conditional paths}.
Note that the last condition does not require the first visit, i.e., after $i=n$, the multiple visits to $x=x^*$ are allowed. In Fig. ~\ref{figS2}, we show sample averages of such conditional paths with $x^* = 0$ for (a) sub-diffusive and (b) super-diffusive fBms.  Although plotted paths look quite irregular due to the limited number of sample paths, one can conceive a clear tendency that (a) sub-diffusive conditional paths are pulled back to the physical domain $x>0$ after touching $x=0$, (b) super-diffusive conditional paths are prone to keep the moving direction.

\subsection{Average heat in the post-first-passage process}

Average heat in the post-first-passage process is plotted in Fig.~\ref{fig4} in the main text for the cases of subdiffusive fBm $\alpha=0.8$ and superdiffusive fBm $\alpha=1.2$. Here, we plot the same quantities for the cases of $\alpha=0.5$ and $\alpha=1.5$ in Fig.~\ref{figS3}.

\section{Classification of FTs by entropic functionals}
\label{Class_FT}

According to the classification of the path probability ratios by entropic functionals~\cite{AdvPhys_Roldan_2023}, 
$\Sigma$-stochastic entropic functional is defined to take a pair of the path probabilities of the original with that of the temporal reverse while $\Lambda$-stochastic entropic functional may take an arbitrary pair of path probabilities.
The conventional fluctuation theorem usually begins with introducing (i) the temporal pair mentioned in the $\Sigma$-stochastic entropic functional.
Since the importance of a pair of probabilities was realized, different types of fluctuation theorems have been discovered.
As remarkable forms, besides (i), there are three kinds of the FTs under nonequilibrium steady state~\cite{AdvPhys_Roldan_2023,PRL_Hatano_Sasa_2001,JStatMech_Chernyak_Chertkov_Jarzynski_2006,JPhysAMathGen_Speck_Seifert_2005,JStatMech_GarciaGarcia_2012}, where (ii) extra heat is extracted by constructing a ratio of the path probability of the original with that of the temporal reverse with the probability flow maintained~\cite{PRL_Hatano_Sasa_2001,JStatMech_Chernyak_Chertkov_Jarzynski_2006,JStatMech_GarciaGarcia_2012}, (iii) house-keeping heat is observed from a path probability ratio of the original probability flow with that of the temporal reverse flow with the paths kept unchanged~\cite{JPhysAMathGen_Speck_Seifert_2005,JStatMech_Chernyak_Chertkov_Jarzynski_2006,JStatMech_GarciaGarcia_2012}.
The last one (iii) with the house-keeping heat is classified into $\Lambda$-stochastic entropic functionals.
The analogous FT relation in the MIs is distinct from those, but, when applied into this classification, a pair of the MI paths made by the spatial reflection creates a notable subclass in the $\Lambda$-stochastic entropic functionals.

%%===========================================================================================%%
%% If you are submitting to one of the Nature Portfolio journals, using the eJP submission   %%
%% system, please include the references within the manuscript file itself. You may do this  %%
%% by copying the reference list from your .bbl file, paste it into the main manuscript .tex %%
%% file, and delete the associated \verb+\bibliography+ commands.                            %%
%%===========================================================================================%%

%% \bibliography{sn-bibliography}% common bib file
%% if required, the content of .bbl file can be included here once bbl is generated
%%\input sn-article.bbl

\end{document}